\documentclass{emulateapj}

\newcommand{\msun}{$M_{\odot}$}

\newcommand {\HI}      {\ion{H}{1}}      %  HI
\newcommand {\HII}     {\ion{H}{2}}      %  HII
\newcommand {\HeI}   {\ion{He}{1}}   %  HeI
\newcommand {\HeII}   {\ion{He}{2}}   %  HeII

\newcommand {\kms}    {km~s$^{-1}$}

\newcommand {\etal}   {et~al.}

\begin{document}

\title{The Efficiency of Stellar Reionization:  \\
Effects of Rotation, Metallicity, and Initial Mass Function}

\author{Michael W. Topping\altaffilmark{1} and J. Michael Shull\altaffilmark{2}}
\affil{CASA, Department of Astrophysical \& Planetary Sciences, \\
University of Colorado, Boulder, CO 80309 \\
mtopping@astro.ucla.edu, michael.shull@colorado.edu}  

\altaffiltext{1}{now at Department of Physics \& Astronomy, UCLA, Los Angeles, CA  90095-1547}  
\altaffiltext{2}{also at Institute of Astronomy, Cambridge University, Cambridge CB3~OHA, UK} 

%****************************************************************** %
\newpage
\begin{abstract} 

We compute the production rate of photons in the ionizing Lyman continua (LyC) of \HI\ 
($\lambda \leq 912$~\AA), \HeI\ ($\lambda \leq 504$~\AA), and \HeII\ ($\lambda \leq 228$~\AA) 
using recent stellar evolutionary tracks coupled to a grid of non-LTE, line-blanketed \texttt{WM-basic}
model atmospheres.  The median LyC production efficiency is $Q_{\rm LyC} =  (6\pm2) \times 10^{60}$ 
LyC photons per \msun\ of star formation (range $[3.1-9.4] \times 10^{60}$) corresponding to a
revised calibration of $10^{53.3\pm0.2}$ photons~s$^{-1}$ per \msun~yr$^{-1}$.
Efficiencies in the helium continua are 
$Q_{\rm HeI} \approx 10^{60}$ photons $M_{\odot}^{-1}$ and  $Q_{\rm HeII} \approx 10^{56}$ photons 
$M_{\odot}^{-1}$ at solar metallicity and larger at low metallicity.  The critical star formation rate 
needed to maintain reionization against recombinations at  
$z \approx 7$ is $\dot{\rho}_{\rm SFR} = (0.012\; M_{\odot}\; {\rm yr}^{-1}\; {\rm Mpc}^{-3}) 
[(1+z) / 8]^3 (C_{\rm H} / 3) (0.2/ f_{\rm esc})$ for fiducial values of IGM clumping factor 
$C_H \approx 3$ and LyC escape fraction $f_{\rm esc} \approx 0.2$.  The boost in LyC production 
efficiency is an important ingredient, together with metallicity, $C_H$, and $f_{\rm esc}$, in assessing 
whether IGM reionization was complete by $z \approx 7$.  Monte-Carlo sampled spectra of coeval 
starbursts during the first 5 Myr have intrinsic flux ratios of $F(1500)/F(900) \approx 0.4-0.5$ and 
$F(912^-)/F(912^+) \approx 0.4-0.7$ in the far-UV (1500~\AA), the LyC (900~\AA), and at the 
Lyman edge (912~\AA). These ratios can be used to calibrate the LyC escape fractions in starbursts.   

\end{abstract} 


\clearpage

%%%%%%%%%%%%%%%%%%%%%%%%%%%%%%%%%%%%%%%%%%%%%%%%%%%%%%%%%  

\section{INTRODUCTION}

Photons in the Lyman continuum (LyC) of hydrogen (energies $E \geq 13.60$~eV and wavelengths 
$\lambda \leq 911.753$~\AA) play a crucial role in theoretical and observational studies of \HII\ regions, active galactic nuclei 
(AGN), starburst galaxies, and ionization of the interstellar medium (ISM) and intergalactic medium (IGM).   The dominant 
sources of ionizing photons in the ISM are O-type stars (Spitzer 1978), while the metagalactic ionizing background
arises from both O-type stars and AGN (Haardt \& Madau 2001, 2012; Shull \etal\ 1999).  Photoionization by hot stars is also 
the likely process governing the \HI\ epoch of reionization at redshifts $z \approx 7$) ending the transition from the ``dark ages" 
following cosmological recombination at $z \approx 1100$ through ``first light" at $z \approx 30-50$.  

The redshift frontier for detection of early galaxies is now at $z \approx 10$ using the Lyman-break flux dropout technique 
with the Wide-Field Camera-3 (WFC3) aboard the {\it Hubble Space Telescope} (Bouwens \etal\ 2011; Bradley \etal\ 2012;
Ellis \etal\ 2013; Robertson \etal\ 2013; Schmidt \etal\ 2014).  These distant galaxies allow us to explore stellar populations 
during the epoch ($z = 7-10$) when reionization was well underway.  The critical star formation rate (SFR) density,
$\dot{\rho}_{\rm SFR} \approx 0.02\ \,M_{\odot}~{\rm yr}^{-1}~{\rm Mpc}^{-3}$ at $z \approx 7$, defines the photoionization
rate needed to balance IGM recombinations (Madau \etal\ 1999;  Shull \etal\ 2012).  However, estimating the number of 
ionizing photons in the radiation field requires calibrating  the O-star production rates and extrapolating the galaxy 
luminosity function well below the observed populations (Trenti \etal\ 2011; Finkelstein \etal\ 2012; Robertson \etal\ 2013).   
The main parameters in $\dot{\rho}_{\rm SFR}$ include the net production rate and escape fraction of Lyman continuum 
(LyC) photons by hot stars and structural properties of the ISM and IGM.  Current estimates suggest that galaxies should 
be able to complete reionization by $z \approx 7$, although the details depend on LyC production efficiencies.

 The LyC production efficiency (ionizing photons per solar mass of star formation) is a standard parameter in cosmological 
 simulations of galaxy formation and IGM reionization.  Historically, many groups have assumed a constant rate of
 $10^{53.1}$ photons~s$^{-1}$  per $M_{\odot}$~yr$^{-1}$.  The current calculation explores the dependences of this 
 parameter on metallicity, initial mass function (IMF), and stellar rotation.
Because of the importance of O-star ionizing photon production (Vacca \etal\ 1996;  Schaerer \& Vacca 1998; Oey \etal\ 2000;
Kewley \etal\ 2001;  Smith \etal\ 2002;  Sternberg \etal\ 2003; Martins \etal\ 2005;  Leitherer \etal\ 2010) to the study of starburst 
galaxies and the epoch of reionization, there is a need for more accurate calculations.   This can now be done, with improvements
in stellar atmospheres modeling, including three-dimensional expanding atmospheres, non-LTE radiative transfer, 
populations of the hydrogen and helium ground states, and an increased number of spectral lines in the line blanketing and 
force modeling of stellar winds.   For stellar populations, the LyC production rates depend on the initial mass function (IMF), 
evolutionary tracks, and model atmospheres, all of which may change at low metallicity and with rapid stellar rotation 
(Schaerer 2003;  Ekstr\"{o}m \etal\ 2012; Georgy \etal\ 2013).  Some of the recent tracks also include the effects of rotation 
on stellar structure and evolution at different metallicities.   

In this paper, we compute the production efficiency (LyC photons per solar mass of star formation).  We use new evolutionary 
tracks of various metallicities, with and without rotation, together with ionizing spectra from model atmospheres.  We find the 
production rates of photons in the ionizing continua of \HI, \HeI, and \HeII\ by integrating over a grid of stellar tracks coupled 
to non-LTE, line-blanketed stellar atmospheres using the \texttt{WM-basic} code (Pauldrach \etal\ 2001; Puls \etal\ 2005) 
with a consistent metallicity.  Section~2 describes our procedure for deriving the stellar LyC from evolutionary tracks and 
consistent model atmospheres.  Section~3 explains our methods for generating a grid of atmosphere models overlaid on 
evolutionary tracks of solar and sub-solar metallicities, with and without rotation.  In Section~4 we present and analyze 
results for the ionizing photon luminosities and lifetime-integrated LyC production efficiencies, for both the hydrogen LyC 
as well as the continua of \HeI\ and \HeII\ (the latter is especially sensitive to metallicity and rotation).   In Section~4.3, we 
apply the new LyC efficiencies to IGM reionization and critical SFR at $z \approx 7$.   Section~5 provides a summary and 
a discussion of future directions in the field.

\section{COMPUTING THE STELLAR LYMAN CONTINUUM} 

Our LyC calculation relies on coupling a full grid of stellar evolutionary tracks with model atmospheres.  Although it would 
be desirable to explore a wider range of other parameters (mass-loss rates, clumpy and porous winds, and binaries) there 
are no complete evolutionary tracks for such parameters.  Thus, we rely on stellar grids from the Geneva group (Schaerer 
2003; Ekstr\"om \etal\  2012; Georgy \etal\  2013).   Our computational procedure begins with these evolutionary tracks, 
which provide the key stellar parameters:  effective temperature ($T_{\rm eff}$), surface gravity ($\log g$), mass ($M$), and 
radius ($R_*$).   From atomic data and stellar opacities, we run new model atmospheres to compute the spectral energy 
distribution (SED) with astrophysical flux ($F_{\lambda}$) in units of erg~cm$^{-2}$~s$^{-1}$~\AA$^{-1}$.  We derive the 
integrated ionizing photon fluxes, $q_i$ (phot~cm$^{-2}$~s$^{-1}$), from the star's surface for the \HI, \HeI, and \HeII\ 
continua by integrating the modeled spectra over wavelengths
$\lambda \leq \lambda_{\rm lim}^{(i)}$,
\begin{equation}
      q_i = \int_0^{\lambda^{(i)}_{\rm lim}} \frac{\pi \lambda F_{\lambda}}{hc} \, d\lambda \; .  
\end{equation}  
Here, $hc/\lambda$ is photon energy and  $\lambda_{\rm lim}^{(i)}$ correspond to the \HI, \HeI, and \HeII\ continuum edges 
at $\lambda_{\rm lim} = 911.753$~\AA, 504.259~\AA, and 227.838~\AA\ for indices $i =$ 0, 1, and 2 respectively.

We combine the evolutionary tracks and grid of model atmospheres consistently to obtain an updated grid ($T_{\rm eff}$, 
$\log g$) where each point contains information on photon fluxes instead of the entire synthetic spectrum.   
The star's ionizing {\it photon} luminosities $Q_i = 4 \pi R_*^2 q_i$ (photons s$^{-1}$) are found from the star's surface area 
at each point along the evolutionary track for a star of radius $R_*(t)$ and initial mass $M$.  The lifetime-integrated values of 
$Q_0$, $Q_1$, and $Q_2$ are then derived for the \HI, \HeI, and \HeII\ continua.   Thus, $Q_0^{\rm (tot)}(m)$ gives the total
number of ionizing (LyC) photons produced over a massive star's life, where $m =  M / M_{\odot}$ is in solar-mass units.   
For O-type stars in the range $25 \leq m  \leq 100$, typical values are $Q_0 ^{\rm (tot)} \approx (1-10) \times 10^{63}$ photons.  
These photon production numbers can then be summed over a stellar IMF using one of the common distributions (Salpeter 
1955; Kroupa 2001; Chabrier 2003) to determine the number of ionizing photons produced by a stellar population.  
A similar grid of \texttt{WM-basic} model atmospheres at solar metallicity was computed by Sternberg \etal\ (2003).  Our 
calculations are done with updated stellar parameters appropriate for the new evolutionary tracks. 

Reionization models and cosmological simulations require a relationship between SFR and LyC production rate. This 
calibration is often expressed as the standard ratio (Madau \etal\ 1999) of $10^{53.1}$ photons s$^{-1}$ per 
$M_{\odot}$~yr$^{-1}$.  We adopt an average LyC production rate over the age of an OB starburst.  Without great loss 
of accuracy, one can simply cancel the time units ($3.1556 \times 10^7$ s/yr) and quote the equivalent standard ratio as
$Q_{\rm LyC} = 4 \times 10^{60}$  LyC photons per solar mass of star formation.    In this paper, we follow the 
convention used by Shull \etal\ (2012) in their reionization models, where they define the ratio
\begin{equation}
   Q_{\rm LyC} = \frac{\int_{m_{min}}^{m_{max}} Q_i^{\rm (tot)}(m) \xi(m) \, \mathrm{d}m}
       {\int_{m_{min}}^{m_{max}} \xi(m) m  \, \mathrm{d}m}\; .
\end{equation}
Here, $\xi(m)$ represents the differential distribution of stars per unit mass (not per $\log m$ as sometimes defined),
and the integration is between mass limits $m_{\rm min}$ and $m_{\rm max}$.  For convenient units, we express 
$Q_{\rm LyC}$ in $10^{63}$ photons per $M_{\odot}$ of star formation\footnote{The units of $10^{63}$ photons were 
chosen (Shull \etal\ 2012) to match the typical lifetime LyC photon production for O stars with mass $25 \leq m \leq 100$. 
Integration over a stellar IMF reduces $Q_{\rm LyC}$ to much lower values, $0.002 - 0.004$ since the low-mass stars 
contribute mass but essentially no LyC photons.}.
For a range of metallicities, $0.1 \leq (Z / Z_{\odot}) \leq 1$, older model atmospheres, and a Salpeter IMF, 
Shull \etal\ (2012) found that $Q_{\rm LyC} \approx 0.0024-0.0038$ ($\times10^{63}$ photons per $M_{\odot}$).   
Below, we update this calibration factor using new stellar data on LyC production and exploring the effects of lower 
metallicity and increased rotation.

\section{COMPUTATIONAL METHODS FOR TRACKS AND ATMOSPHERES}

The earliest stellar evolutionary tracks included only a small amount of stellar physics that was computationally allowed at the 
time.  These models solved the equations of stellar structure with little treatment of more advanced physics.  Some of the 
first standard evolutionary tracks came from the Geneva group in the early 1990's (Schaller \etal\ 1992;  Schaerer \etal\ 1993a,b;
Charbonnel \etal\ 1993).  After observations suggested higher mass loss rates for massive stars, additional set of tracks were 
produced (Meynet \etal\ 1994; Fagotto \etal\ 1994a,b).  The next major advance was to add the effects of stellar rotation. 
Although the early efforts were productive, they only described some effects of rotation and did not extend to
$M < 9\;M_{\odot}$ (Meynet \& Maeder 2000).  

For the past two decades, the most widely used stellar evolutionary tracks were those of Schaller \etal\ (1992).  The most recent
tracks (see Figure~1) computed at solar metallicity (Ekstr\"om \etal\ 2012) and sub-solar metallicity (Georgy \etal\ 2013) are 
expected to become the new standard in the field.  These models surpass previous evolutionary tracks with their treatment of 
mass loss from the stellar wind, more accurate nuclear burning rates, and the addition of stellar rotation. Advances in 
helioseismology and other stellar observations have improved our understanding of the interior properties of rotating stars.  
From these data, Ekstr\"om \etal\ (2012) created evolutionary tracks and isochrones for a wide mass range of rotating and 
corresponding non-rotating stars.   These models can be used to create an initial picture of rotating stellar evolution. One of the 
most important differences with rotation is mixing caused by meridional circulation, which continuously provides new fuel to 
burn in the core and increases the lifetime of a star.

%%%%%%%%%%%%   Figure 1   %%%%%%%%%%%%%%%%%%%%%%%%%%%%%%%%%%%%%%%%%%

\begin{figure}
    \epsscale{1.2}
    \plotone{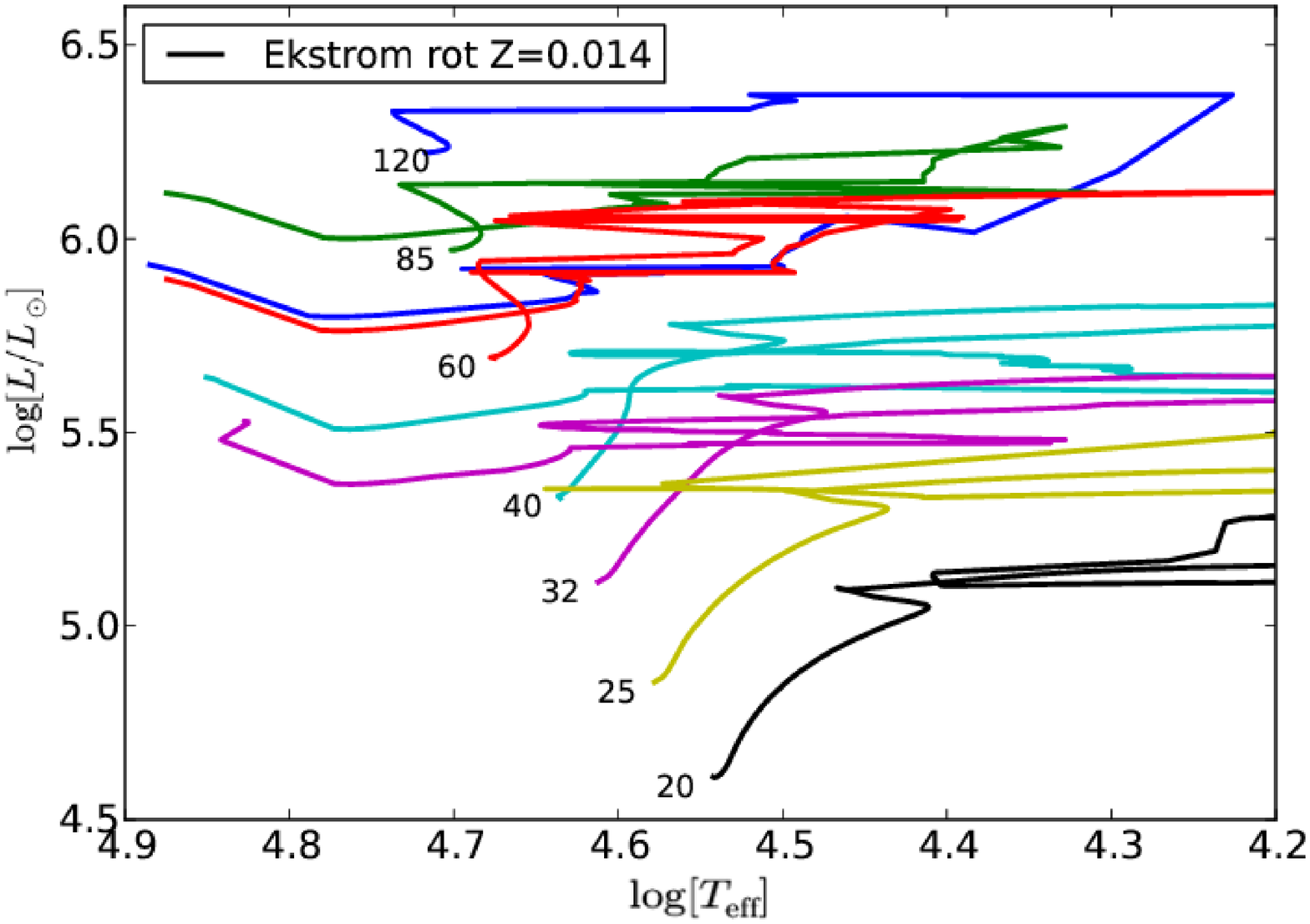}
    \plotone{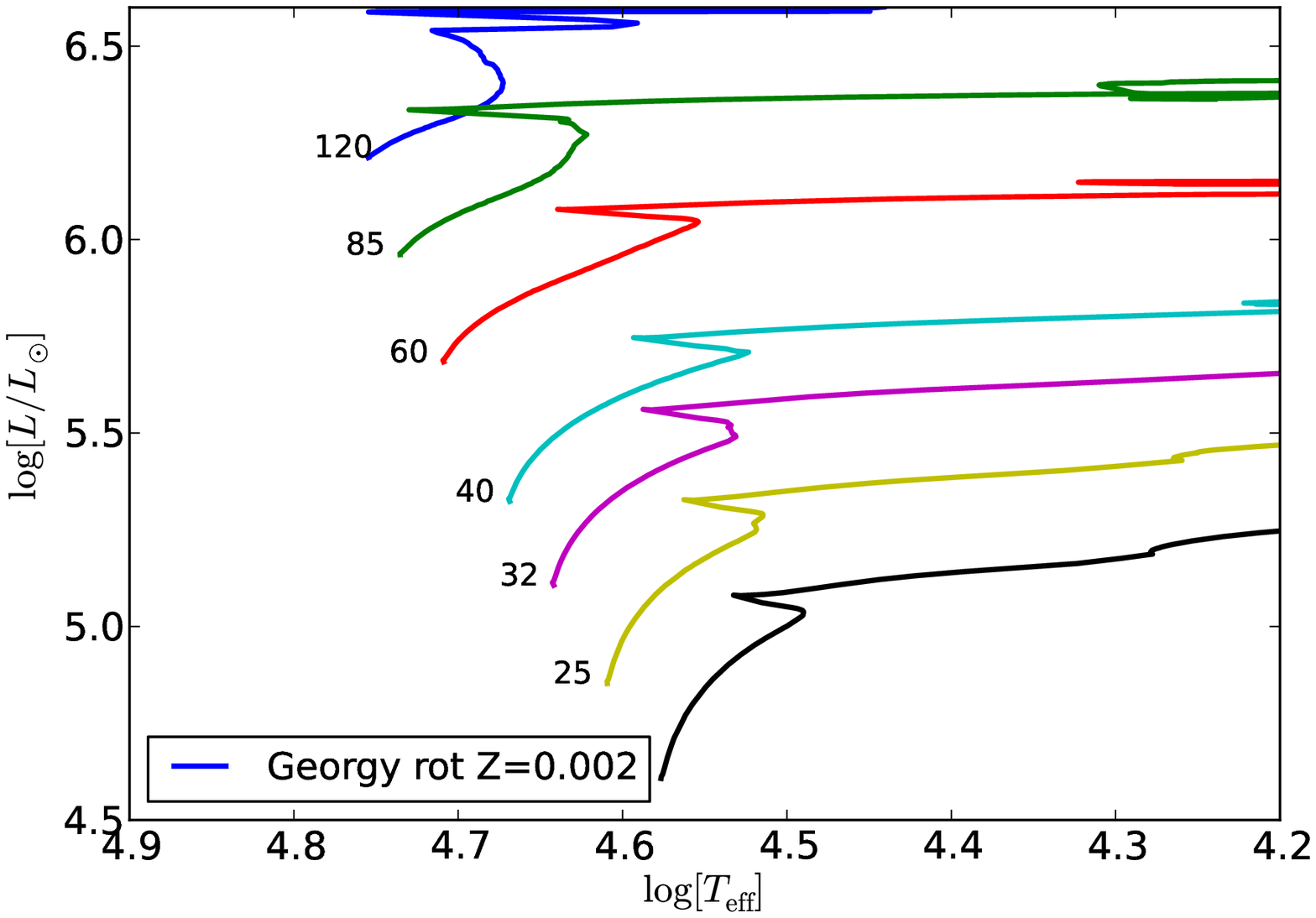}
    \caption{New rotating tracks (Ekstr\"om \etal\ 2012; Georgy \etal\ 2013).  Tracks computed at solar metallicity (top)
   and sub-solar metallicity (bottom).  Note different post main sequence populations at low metallicity, with fewer
     ``blue-loop"  excursions and fewer WR stars. }
\end{figure}

 %%%%%%%%%%%%%%%%%%%%%%%%%%%%%%%%%%%%%%%%%%%%%%%%%%%%%%%%%%%%

Improvements in the quality of model atmospheres include non-LTE treatment of the radiation field, wind and line blanketing, 
and increased code efficiency.  The first upgrades involved an efficient method for approximating the radiation pressure due to 
spectral lines.  The CAK model (Castor, Abbott \& Klein 1975) averaged the radiative force over all lines instead of treating them 
separately.  This increase in radiative force led to higher mass loss rates.  Even with this advance, atmosphere models were 
restricted to local thermodynamic equilibrium (LTE), requiring a blackbody radiation field and Boltzmann level populations.   
Pauldrach \etal\ (1986) upgraded the CAK model by neglecting the radial-streaming approximation for photons that drive the 
stellar wind.  Rotational effects were found to be negligible for rotational velocities $v_{\rm rot}  < 200$ km s$^{-1}$. 

Advances in model atmospheres have been more consistent than advances in evolutionary tracks, primarily because of the 
availability of fast computers.  The most efficient codes still use a modified version of the CAK method to compute line forces, 
because it is fast and its approximations provide efficiency.  Perhaps the most important improvement is the extension of the 
models into the non-LTE regime, allowing the radiation field to deviate from Planckian and level populations from Boltzmann.   
Other advances include solving the radiative transfer equations in co-moving coordinates, moving away from plane-parallel 
models, and increasing the accuracy of atomic data.  With these recent improvements, a model atmosphere can now be 
computed in less than ten minutes, allowing thousands of models to be computed over the course of the project and providing 
accurate time resolution for the evolution of stellar  spectra.   A few of the most prominent atmosphere codes are:
\texttt{TLUSTY} (Lanz \& Hubeny 2003, 2007), \texttt{FASTWIND} (Puls \etal\ 2005), \texttt{CMFGEN} (Hillier \& Miller 1998), 
and \texttt{WM-basic} (Pauldrach \etal\ 2001).  

For our model atmospheres we chose the \texttt{WM-basic} 
code\footnote{Code can be found at http://www.usm.uni-muenchen.de/people/adi/Programs/Programs.html} primarily 
because of its hydrodynamic solution of spherically expanding atmospheres with line blanketing and non-LTE radiative 
transfer and its treatment of the continuum and lines in the ionizing EUV.   This code is also a standard for theoretical 
atmosphere modeling in the population synthesis code \texttt{Starburst99} (Leitherer \etal\ 1999, 2014). 
An extensive discussion of hot-star atmosphere codes appears in papers by Martins \etal\ (2005) and Leitherer \etal\ 
(2014).  A comparison of the ability of these codes to fit stellar lines, effective temperatures, and surface gravities was made 
by Puls (2008) and Massey \etal\ (2013).  In Figure 2 we show the results  of \texttt{WM-basic} and \texttt{CMFGEN} for three 
cases with $T_{\rm eff} \approx$ 30kK, 40kK, and 50kK.  The agreement is reasonably good, given the different model
input parameters required to arrive at similar effective temperatures, but slightly different $\log g$ values.

%%%%%%%%%%%%   Figure 2 a,b,c  %%%%%%%%%%%%%%%%%%%%%%%%%%%%%%%%%%%%%%%%%%

\begin{figure*}
    \epsscale{0.65}
    \plotone{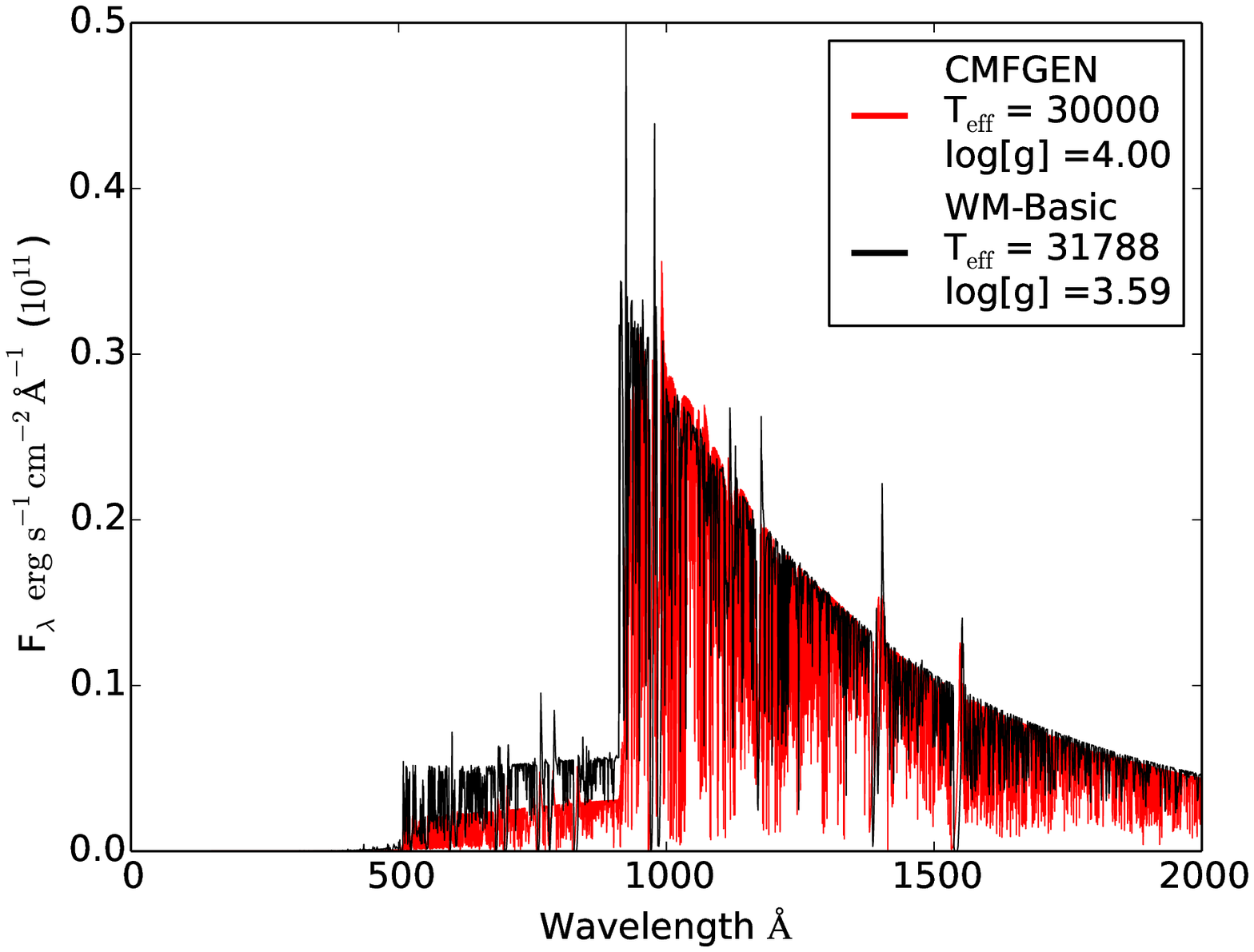}
    \plotone{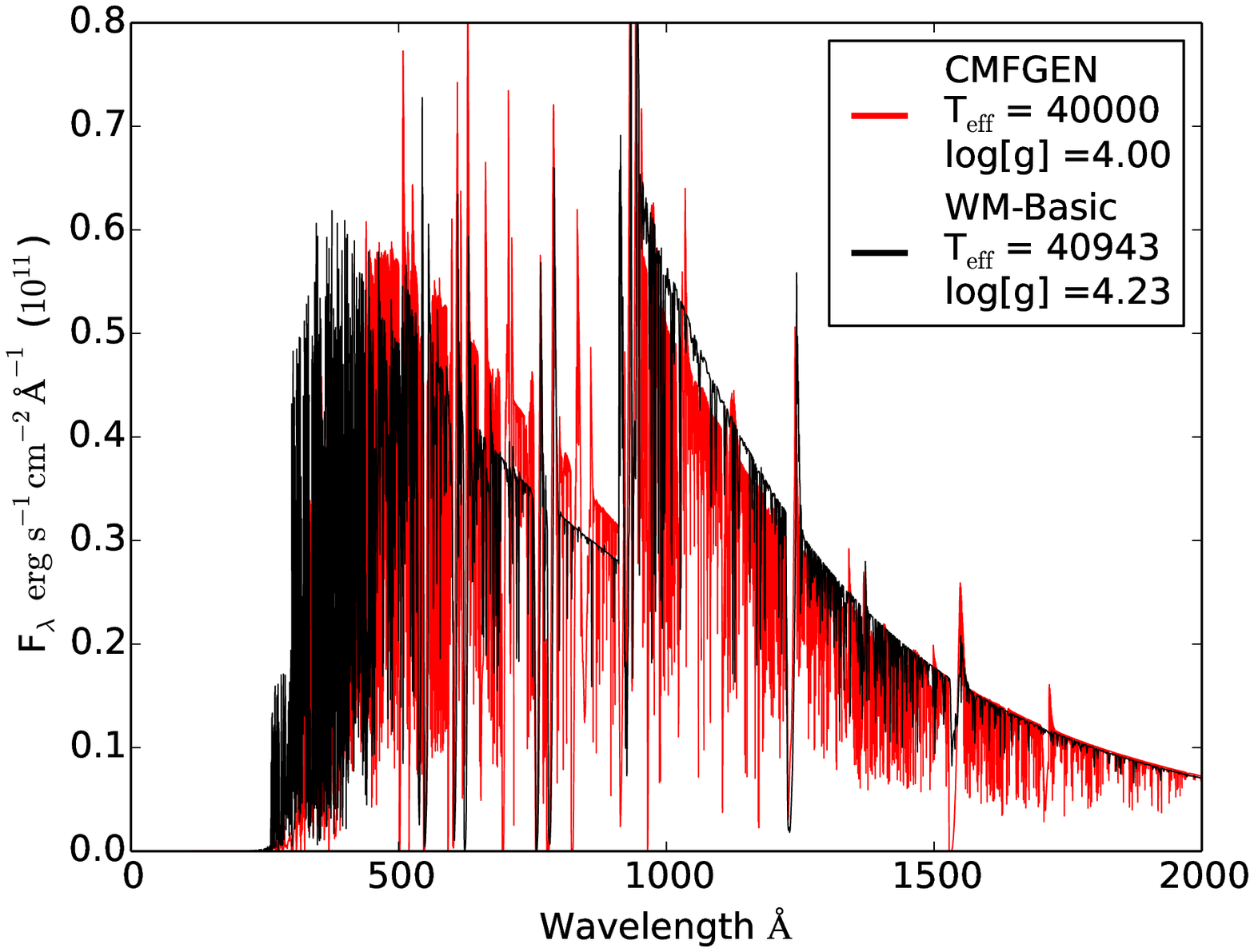}
    \plotone{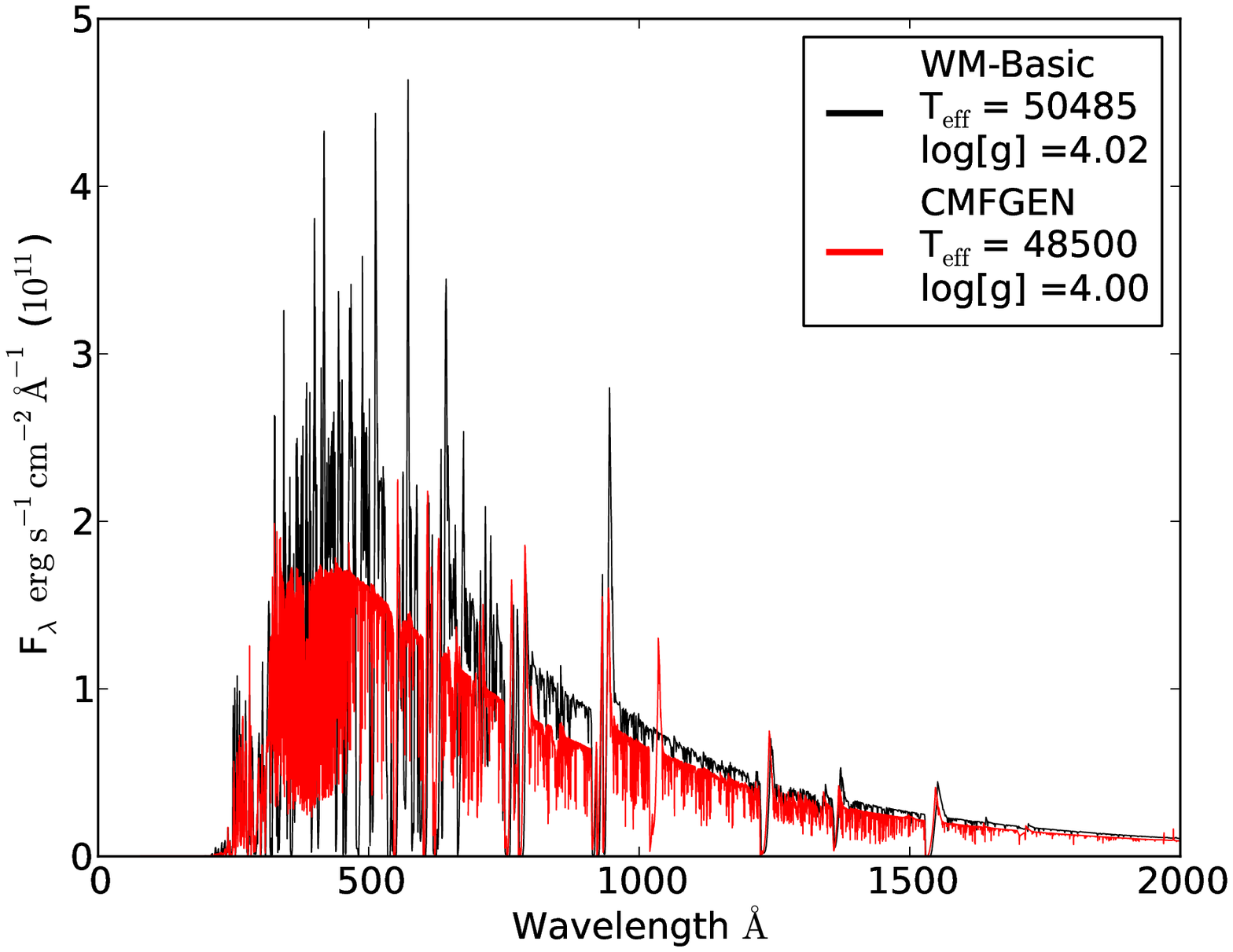}
\caption{Overlaid comparison of model atmospheres at effective temperatures $T_{\rm eff} \approx 30$ kK, 40~kK, 
    and 50~kK, computed using the \texttt{WM-basic} and \texttt{CMFGEN} code with emergent fluxes $F_{\lambda}$ 
    in units $10^{11}$ erg~cm$^{-2}$ s$^{-1}$~\AA$^{-1}$.  Because of slight differences in mass-loss rates and
    required input parameters for the stellar models, we are unable to precisely match effective temperatures ($T_{\rm eff}$) 
    and surface gravities ($\log g$). The \texttt{WM-basic} model is shown in black and the \texttt{CMFGEN} model in red.  }
\end{figure*}

%%%%%%%%%%%%%%%%%%%%%%%%%%%%%%%%%%%%%%%%%%%%%%%%%%%%%%%%%%%

\subsection{Evolutionary Tracks}

We use evolutionary tracks from Schaller \etal\ (1992), Ekstr\"om \etal\ (2012), and Georgy \etal\ (2013).  These models 
include fixed prescriptions for stellar mass loss during late stages of evolution, and they provide a convenient means of
assessing the effects of metallicity and rotation.  It would be useful to assess the dependence of LyC production rates
on alternative prescriptions for mass loss, particularly during late stages of massive-star evolution.  However, at the current 
time, we are constrained to using the available evolutionary tracks.  The field of massive-star evolution continues to present
challenges to the correct implementation of these effects (Vink 2014; Martins 2014; Hirschi 2014) as well as those of 
binaries and stellar mergers (Eldridge \etal\ 2008; de Mink \etal\ 2013, 2014).  Populations of very massive stars 
(150--300 \msun) may exist in the Arches cluster (Figer \etal\ 2002; Crowther \etal\ 2010), in the dense R136 star 
cluster in 30~Doradus, and in the Galactic cluster NGC~3603 (Crowther \etal\ 2010; Martins 2014).  Figer (2005) used 
the Arches cluster to set an upper mass limit of 150~\msun, owing to an absence of 
stars\footnote{Subsequent studies (Espinoza \etal\ 2009) of stars in the Arches cluster found a classical IMF with mass 
distribution $dN/d(\log m) \propto m^{-1.1\pm0.2}$ (Salpeter index is $\Gamma = 1.35$) for masses $m > 10~M_{\odot}$ 
and measured  through nearly 20 magnitudes of visible extinction.  Crowther \etal\ (2010) examined several of the Arches 
cluster stars and derived larger luminosities owing to a larger assumed distance and different extinction law.  They reported 
a few masses in excess of 160~\msun.} with initial masses $\geq130$~\msun.
These mass estimates are subject to uncertainties in stellar luminosity arising from cluster distance, extinction laws, and
effective temperatures.   As summarized in recent reviews (Vink 2014; Martins 2014), there are several stars that
could have masses between 100-200~$M_{\odot}$.    Such stars could provide additional sources of LyC by extending 
the IMF above the last grid point on the evolutionary tracks.   On the other hand, the most massive stars in dense clusters 
are deeply embedded in star-forming clouds, and their LyC may not escape until after these regions clear of gas and dust.  

For the current calculation, we truncate all IMFs at a maximum mass of 120~$M_{\odot}$, and we use the LyC production 
rates to describe massive star formation regions and reionization of the IGM.  The tracks are computed at five different 
metallicities, of which we consider three for non-rotating stars.   In the Hertzsprung-Russell (H-R) diagram stellar evolution
defines the basic parameters ($M_*$, $L_*$, $T_{\rm eff}$, $\log g$) which are related by self-consistency.  For example, 
surface gravity $g = GM_*/R_*^2$ depends on stellar mass $M_*$ and radius $R_*$, and stellar luminosity 
$L_* = 4\pi R_*^2  \sigma T_{\rm eff}^4$ depends on effective temperature $T_{\rm eff}$ and $R_*$.  The ionizing fluxes are 
sensitive primarily to $T_{\rm eff}$.   We use different evolutionary tracks to investigate how different stellar physics can 
affect the production of LyC photons.   The first set of  tracks are from Schaller \etal\ (1992), a past standard for non-rotating 
stars with sub-solar, solar, and super-solar metallicities, $Z = 0.004$, 0.02 and 0.04, respectively, by mass. 
The second set of tracks comes from the same group (Ekstr\"om \etal\ 2012) with a mass range extended 
to $0.8$ \msun\ $\le$ $m$ $\le$ $120$ \msun\  and finer mass resolution.  One essential advance is their treatment of rotation.  
We use two sets of tracks with metallicities $Z=0.004$ and $Z=0.014$, each computed with an initial rotation of 40\% of the 
critical rotation speed.  We also consider their tracks without rotation.   Since the Schaller \etal\ (1992) work, the accepted 
value for the solar metallicity has decreased to either $Z_{\odot} = 0.014$ (Asplund \etal\ 2009) or $Z_{\odot} = 0.0153$ 
(Caffau \etal\ 2011).   Ekstr\"om \etal\ (2012) calculated both rotating and non-rotating tracks at solar metallicity ($Z=0.014$),
while Georgy \etal\ (2013) computed both rotating and non-rotating tracks at sub-solar metallicity ($Z = 0.002$).  The Geneva 
group has not yet provided new tracks with super-solar metallicity.

\subsection{Model Atmospheres}

After compiling the evolutionary tracks, we produce model spectra at each time step along the track.  We use the atmosphere code 
\texttt{WM-basic} because of its efficiency and ability to generate accurate spectra, including lines in the UV and EUV.  The code requires 
only a few basic parameters ($T_{\rm eff}$, $\log g$, $R_*$, abundances) to compute a model.  For each set of tracks we compute a 
grid of atmospheres in the $(T_{\rm eff}, \log g)$ plane and follow the evolution of the ionizing spectra of a star through time.  We use 
the flux distributions to find the total LyC flux and integrate to calculate the ionizing photon production for each initial mass.  
An atmosphere model can be completed in 10-15 minutes on a desktop computer\footnote{We used an AMD Phenom~II~X6~1075T 
3.0GHz processor and 32Gb RAM.   With the multiple processor cores of this machine, we typically run up to six models at a time.}.  
Over the course of our project,  we ran hundreds of atmosphere models to achieve accurate time resolution along evolutionary tracks. 
One full grid of models for a set of evolutionary tracks contains around 100 atmosphere models.

%%%%%%%%%%%%   Figure 3   %%%%%%%%%%%%%%%%%%%%%%%%%%%%%%%%%%%%%%%%%%

\begin{figure*}
    \epsscale{1.1}
    \plottwo{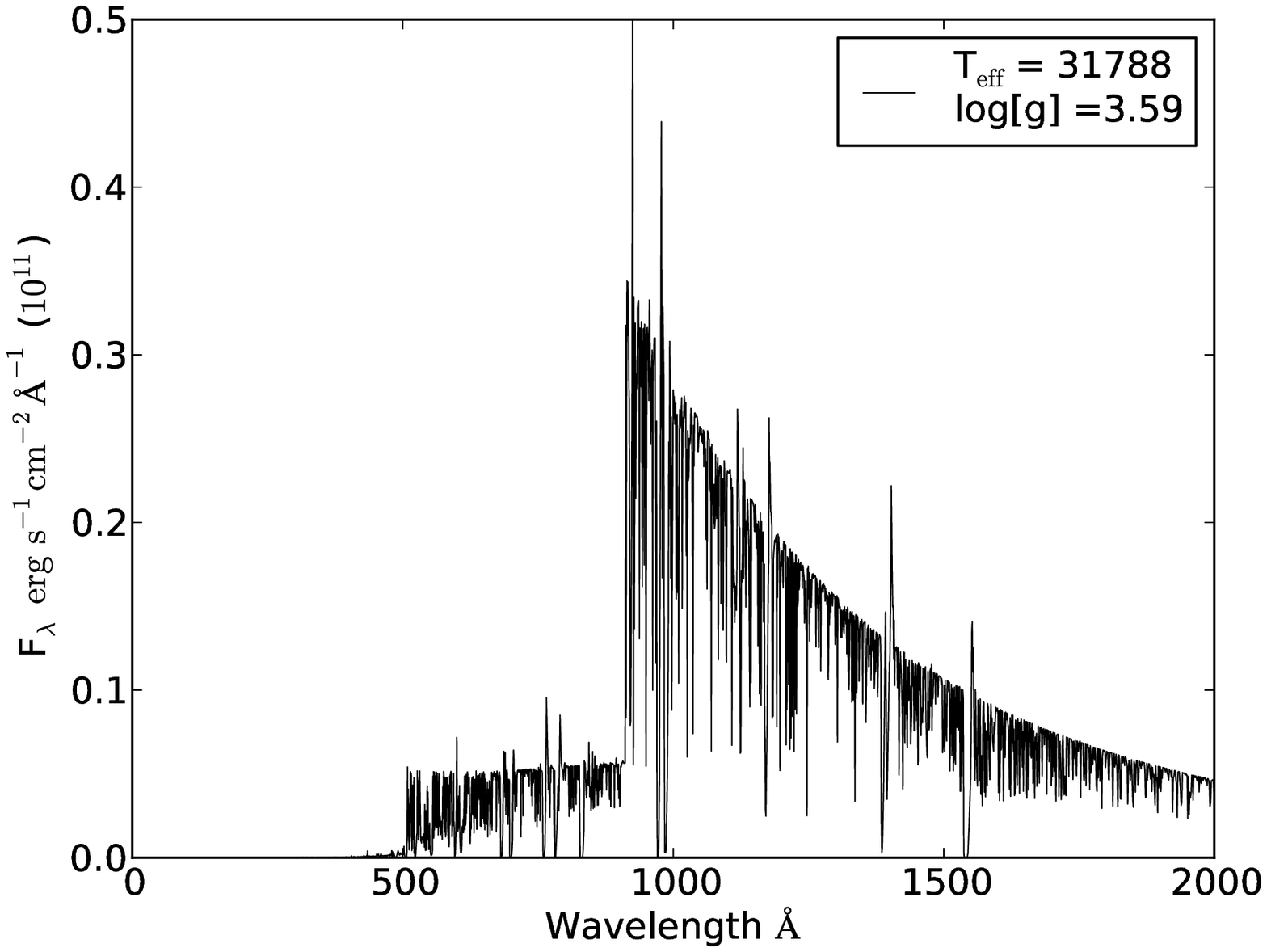}{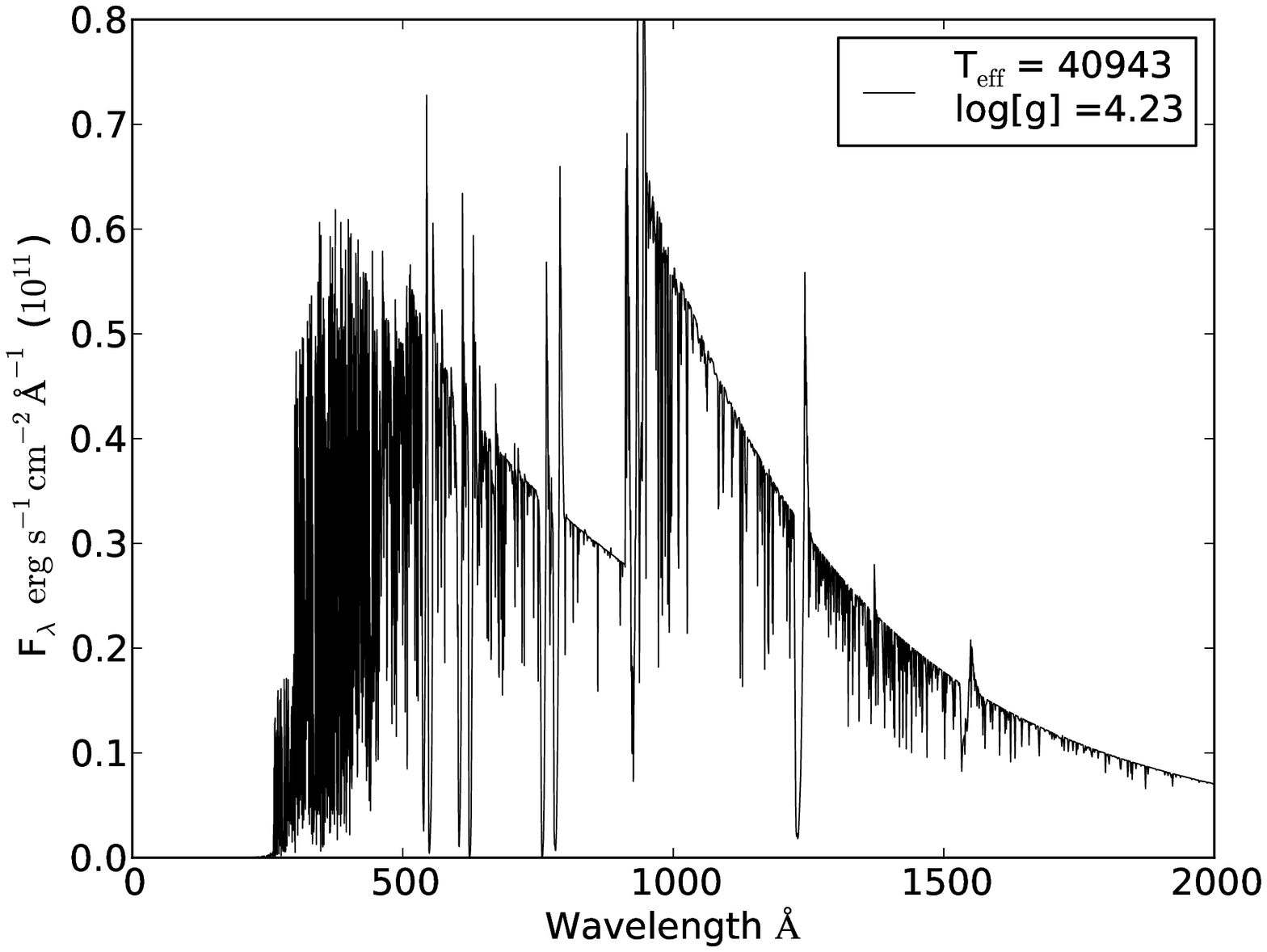}
    \plottwo{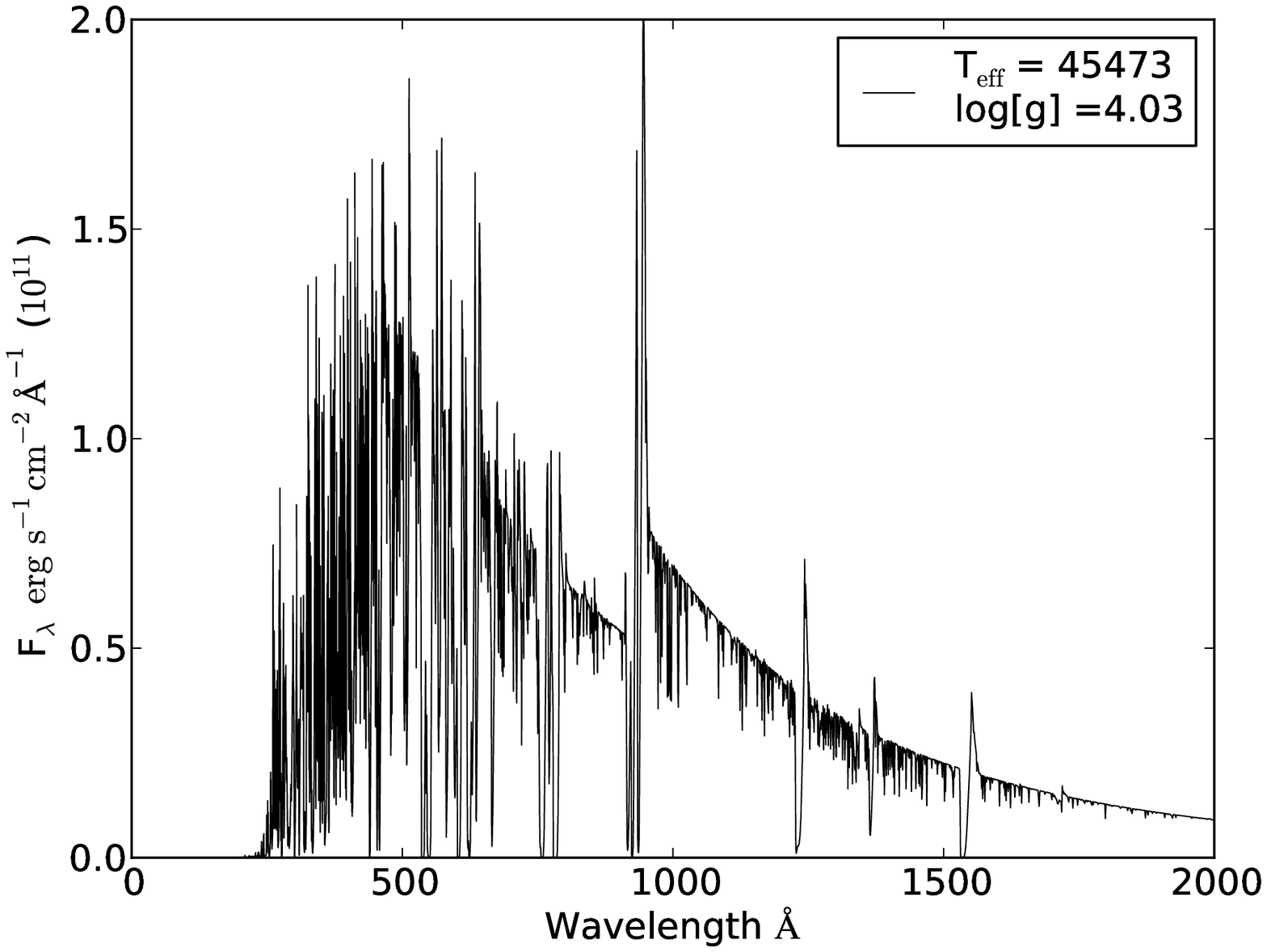}{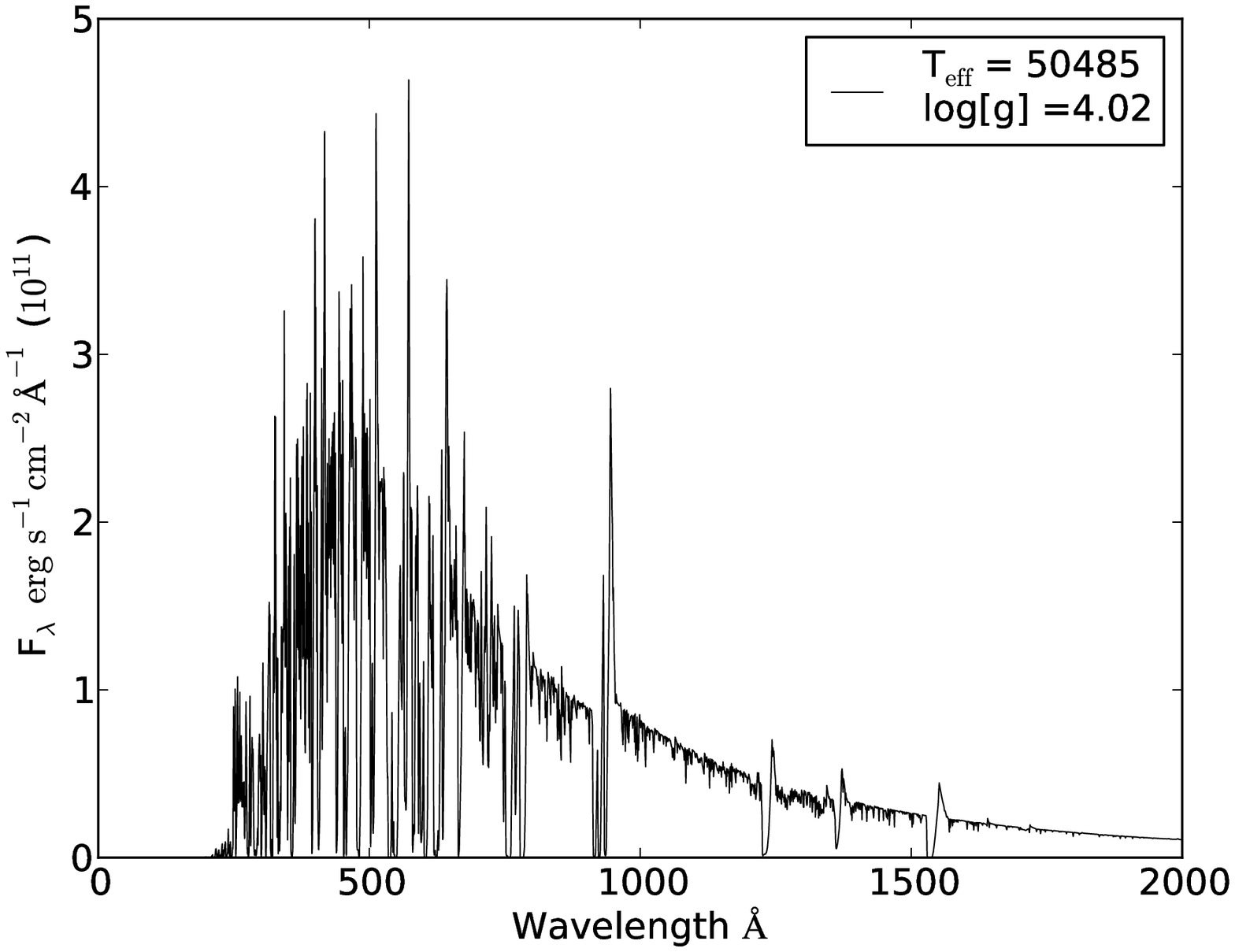}
    \plottwo{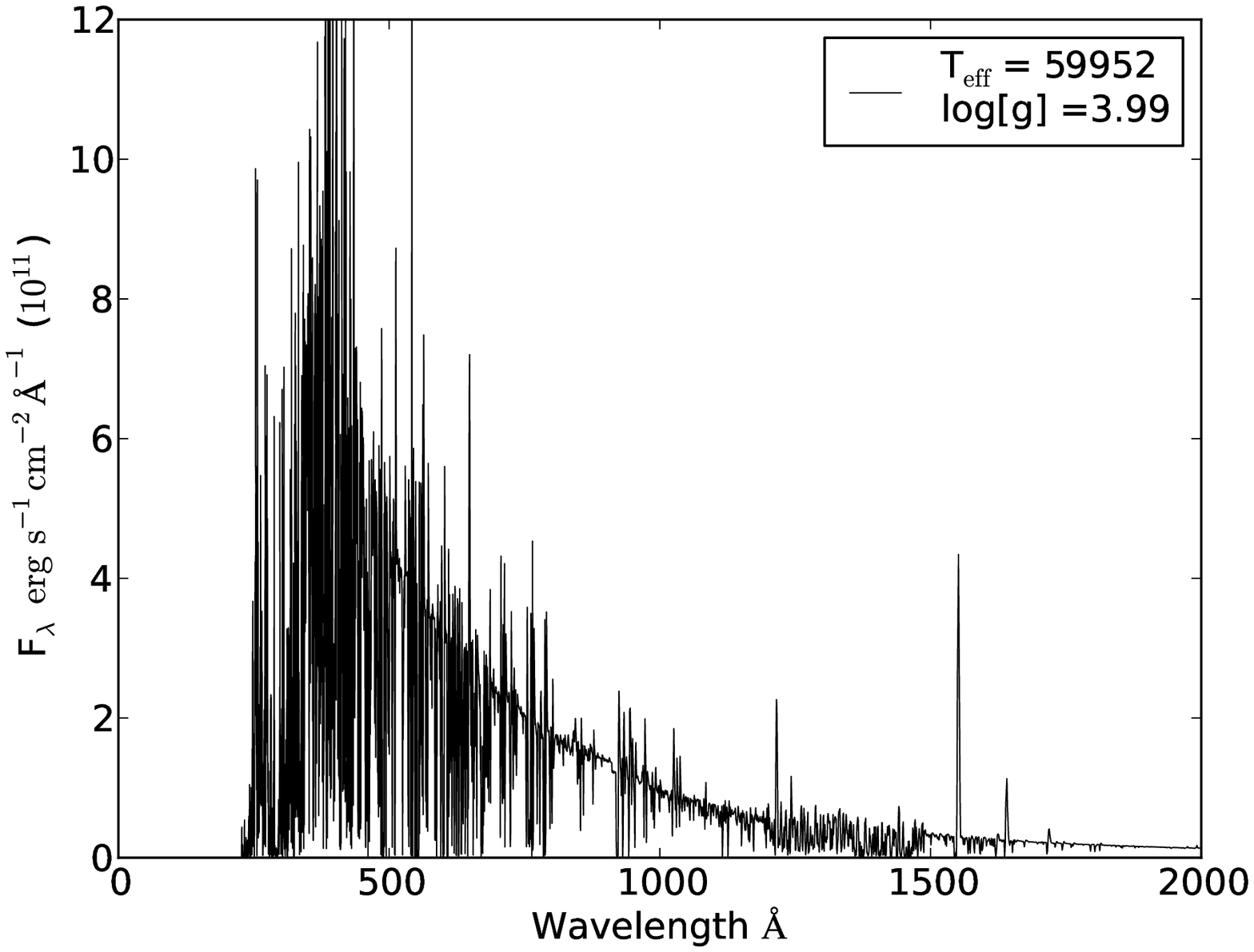}{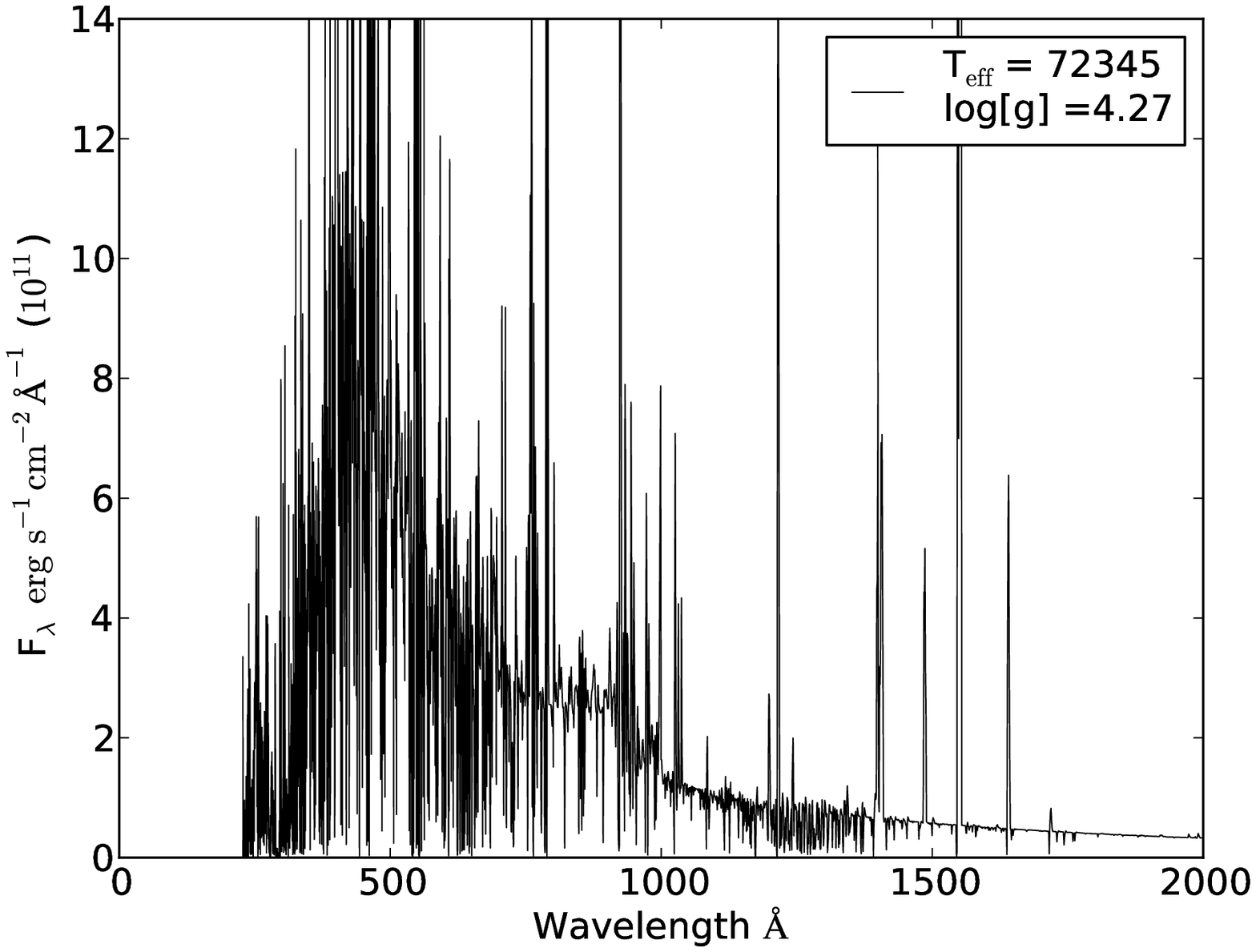}
\caption{Six model atmospheres computed using the \texttt{WM-basic} atmosphere code and taken from 
    the Ekstr\"om \etal\ (2012) grid of rotating tracks at solar metallicity ($Z = 0.014$).  Models are shown in 10kK intervals 
    in $T_{\rm eff}$; we include 45kK because of the rapid change in strength of the H and He edges in that temperature range.  
    Models are all for O-type stars near the main sequence.  \vspace{0.5cm}  }
\end{figure*}

%%%%%%%%%%%%%%%%%%%%%%%%%%%%%%%%%%%%%%%%%%%%%%%%%%%%%%%%%%%

The process of model atmosphere calculation comes in three different pieces: the hydrodynamics, calculation of the occupation numbers 
and radiation field, and production of the synthetic spectrum.  The hydrodynamics section is where the initial parameters are input, most 
importantly $T_{\rm eff}$,  $\log g$, and $R_*$.  From these, the code finds the mass, luminosity, and other basic data.  The next input 
are the abundances of elements between H and Zn, with three options:  solar metallicity, solar metallicity while defining the H and He
 abundances, and custom-defined abundances.  Because the  default ``solar" abundances programmed into the code were outdated, 
 we used the recent standard of $Z_{\odot}=0.014$ by mass (Asplund \etal\ 2009).  The code requires input of the metal abundances
 relative to solar, which we re-examine after a model is run.  In its normalization procedure, one can alter the abundances.  The resulting 
 values may differ from the input values, but generally by only a few percent.  Once the input file is written, the model can be run using the 
 graphical user interface. 

For the radiation force on the atmosphere and wind, we use three line-force parameters ($k$, $\alpha$, $\delta$) in the CAK method.  
This method approximates the sum of a large number of spectral lines, with temperature assumed to be constant throughout the 
atmosphere to increase the efficiency.    The hydrodynamics calculation is iterated two more times.  The first time, the temperature 
gradient is computed using the continuum opacities.  In the last iteration the temperature is computed including the line opacities.  
The code then computes the photon energy, occupation numbers, and radiation field in the NLTE regime.  This step provides the 
radiation field, final temperature structure, information about opacities, and occupation numbers for the various elements using detailed 
atomic models.  The final step in the calculation provides the synthetic spectrum.   Figure 3 illustrates the continuum shapes from 
models with $T_{\rm eff}$ ranging from 31,800~K to 72,300~K, particularly the contributions of emission lines, the depth of the Lyman 
edge, $F(912^-) / F(912^+)$, and the ratio, $F(1500) / F(910)$, of fluxes at 1500~\AA\ and 900~\AA.  The latter two ratios have been 
used to calibrate the escape fraction ($f_{\rm esc}$) of LyC radiation. This observational technique was introduced by Steidel \etal\ 
(2001) and employed by Inoue \etal\ (2005), Shapley \etal\ (2006), and Mostardi \etal\ (2013) to estimate $f_{\rm esc}$ in star-forming 
galaxies at $z = 2.85-3.27$.   In their method, the {\it observed} flux-density ratio, $[F(1500) / F(910)]_{\rm obs}$, is compared to 
estimates of the intrinsic ratios determined from the models.  We return to this issue in Section 4.5, where we discuss 
the intrinsic  values, $[F(1500) / F(910)]_{\rm int} \approx 0.4-0.7$ found in our modeled composite spectra of OB associations with 
coeval starbursts.  

In the models of Figure 3 with $T_{\rm eff} < 50,000$~K, the absorption edge at the Lyman limit is clearly visible at 
$\lambda = 912$ \AA.  Figure~4 shows two models at $T_{\rm eff} \approx 41,000$~K with metallicities $Z = 0.014$ (solar) and 
$0.004$ (sub-solar) by mass.  Each has an intrinsic Lyman edge resulting from the non-LTE level population of the ground state 
of \HI.  Differences in metallicity have a minor effect on the depth of the edge, which arises primarily from surface abundances 
and heating.    However, the EUV fluxes are affected by emission lines in the stellar wind, which increase in strength at higher 
metallicity (left panel of Figure 4).

%%%%%%%%%%%%   Figure 4   %%%%%%%%%%%%%%%%%%%%%%%%%%%%%%%%%%%%%%%%%%

\begin{figure*}
    \epsscale{1.1}
    \plottwo{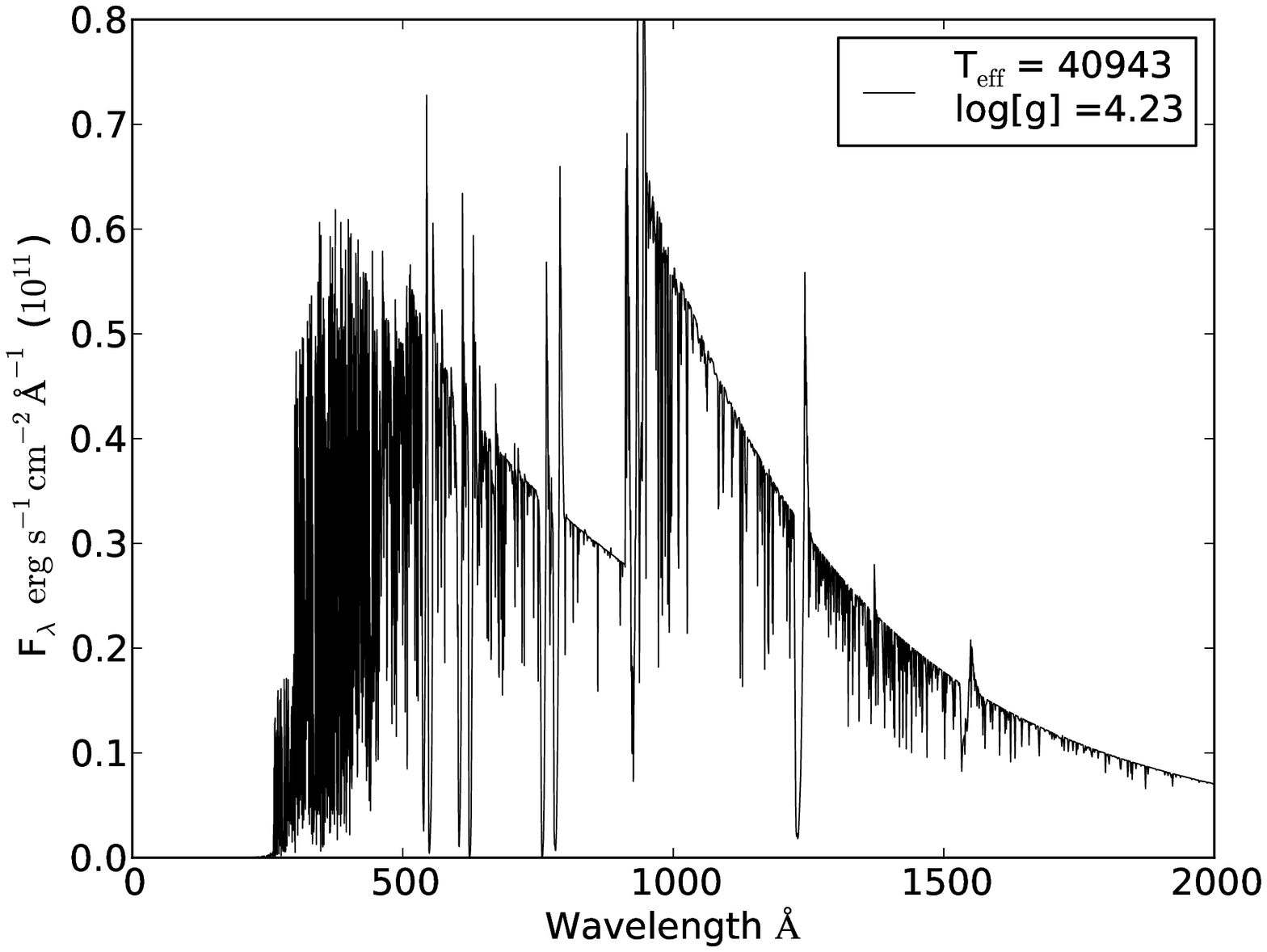}{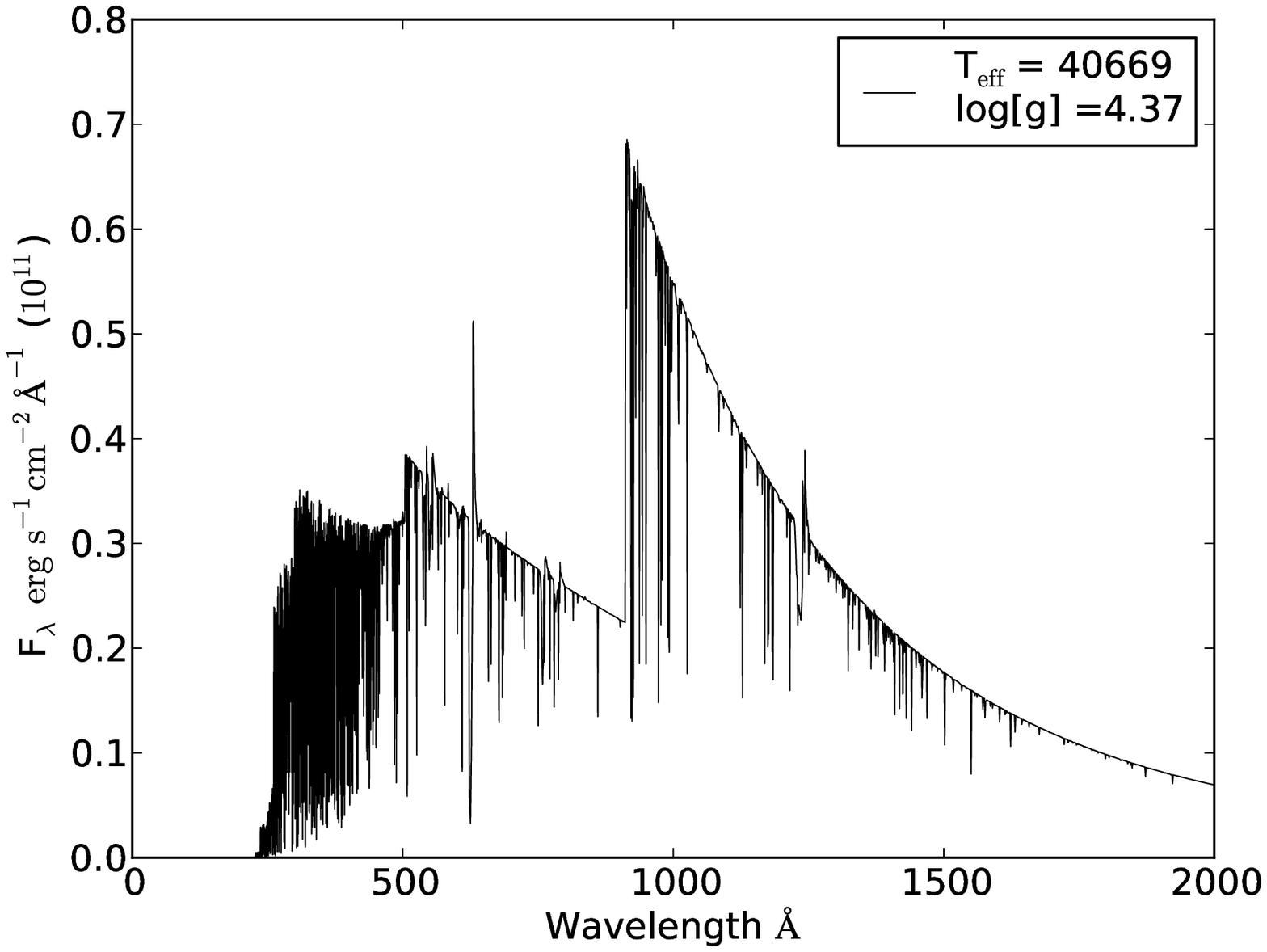}
  \caption{Two model atmospheres with $T_{\rm eff} \approx 41$kK, computed using the \texttt{WM-basic} atmosphere 
     code showing metallicity effects on the strength of the Lyman edge at 912~\AA\ and emission lines
     from the stellar wind.   {\it Left:}  Solar metallicity model ($Z = 0.014$) has a 57\% drop in flux at the continuum 
    edge and substantial metal emission lines in the EUV band (300-900~\AA).   {\it Right:}  Sub-solar metallicity model 
    ($Z = 0.004$) has a 68\% continuum drop and many fewer emission lines. }
\end{figure*}

%%%%%%%%%%%%%%%%%%%%%%%%%%%%%%%%%%%%%%%%%%%%%%%%%%%%%%%%%%%

\subsection{Grid of Atmosphere Models for Evolutionary Tracks} 

In order to find the integrated LyC photon production over an O-star's lifetime, we set up a grid of spectra in the ($T_{\rm eff}, 
 \log g$) plane defined by the initial mass of the evolutionary track (Figure~5).  We distribute the grid points evenly along points 
of equal time.  Because the grid has a higher density where the stars spend more time, we obtain a higher accuracy of photon 
fluxes over the stellar lifetime and improve the accuracy of the integrated photon fluxes.  As the most massive stars shed their 
atmospheres in late stages of evolution (Figure~6), they move quickly through ``blue loops"  and into the Wolf-Rayet (WR) 
phase.  Although the most massive stars spend only a short time in these regions of the H-R diagram, their LyC production 
rate is high. Therefore,  we compute extra atmosphere models and ensure that we place sufficient grid points during these
 late stages to capture the ionizing radiation. 

The WR stars deserve further discussion as the evolutionary descendants of massive stars ($M \geq 25~M_{\odot}$) 
characterized observationally by broad emission lines, strong stellar winds, and mass-loss rates up to 
$10^{-5}~M_{\odot}~{\rm yr}^{-1}$ (Maeder \& Conti 1994; Crowther 2007; Massey 2013). Both luminous ($L > 10^5 \,L_{\odot}$) 
and hot (25,000--100,000~K), WR stars can contribute substantially to the LyC production of a stellar population, particularly in
the \HeI\ and \HeII\ continua.  However, the canonical estimate that WR stars are 10\% of the massive-star population is based 
on little more than the He-burning lifetimes of stars with $M \geq 25~M_{\odot}$ estimated from uncertain evolutionary tracks.  
Because O-star mass-loss rates depend on metallicity, it should be easier to become a WR star at high metallicity.  This 
expectation appears to agree with observations, which show fewer WRs at low metallicity in Local Group galaxies (Maeder \etal\ 
1980; Hamann \etal\ 2006; Neugent \etal\ 2012; Massey \etal\ 2014).   However, the progenitor masses and details of 
post-main-sequence stellar evolution that lead to the various (WC, WN, WO) sub-classes are still poorly understood.  Recent 
surveys of WN and WC stars in the Galaxy, LMC, and M31 (Sander \etal\ 2012; Sander \etal\ 2014; Hainich \etal\ 2014) 
suggest that WN progenitors originate from initial masses between 20--60 \msun.  The ratios of WR stars to O stars (and of 
WN and WC subtypes) are frequently used as tests of massive-star evolutionary models.   Owing to uncertainties in distance 
and reddening and selection biases in detecting WR stars relative to O-type stars, astronomers probably cannot measure the 
Galactic (WR/O) ratio to better than 50\% (Massey \etal\ 1995).

%%%%%%%%%%%%   Figure 5   %%%%%%%%%%%%%%%%%%%%%%%%%%%%%%%%%%%%%%%%%%

\begin{figure}
    \epsscale{1.2}
    \plotone{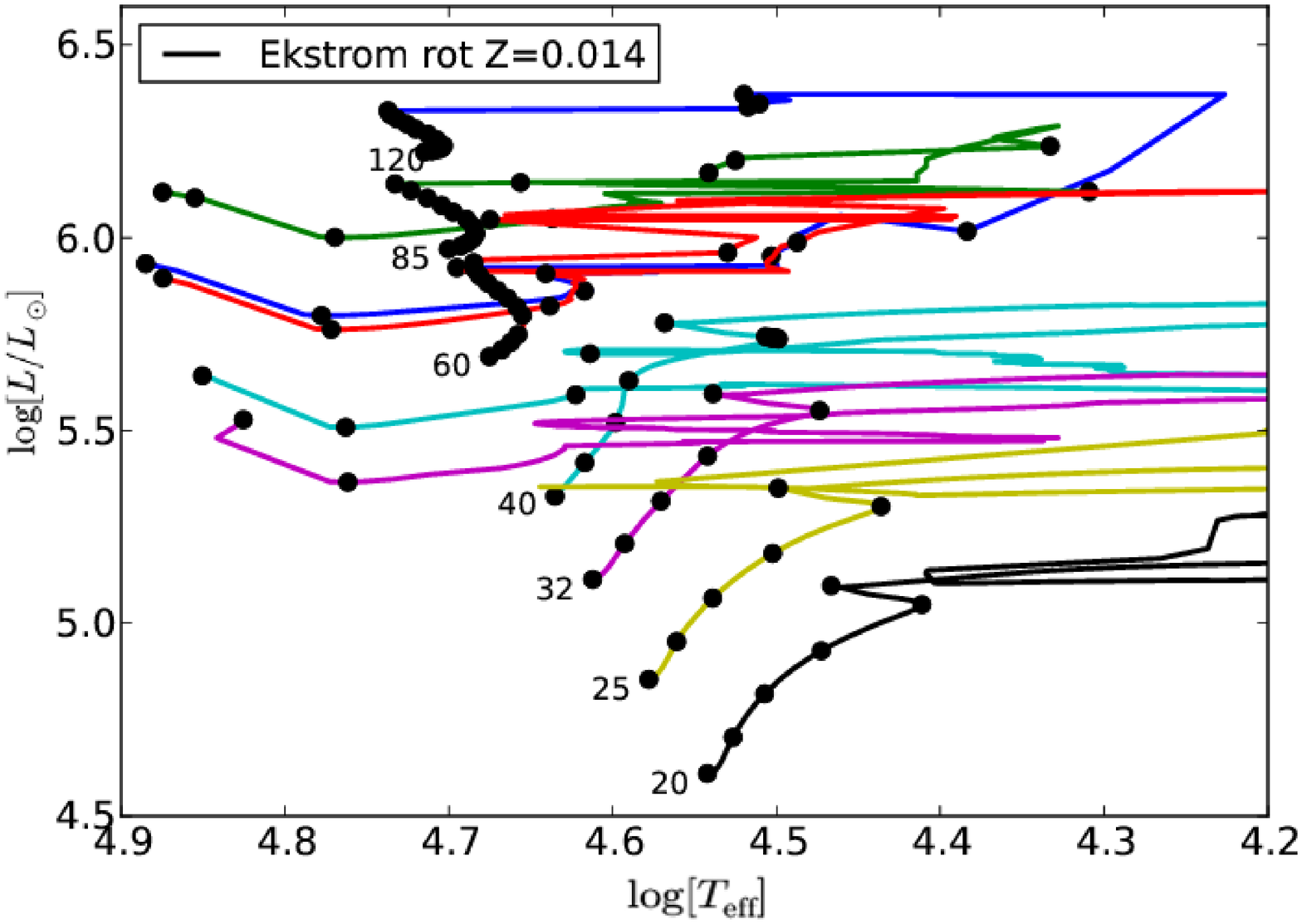}
    \plotone{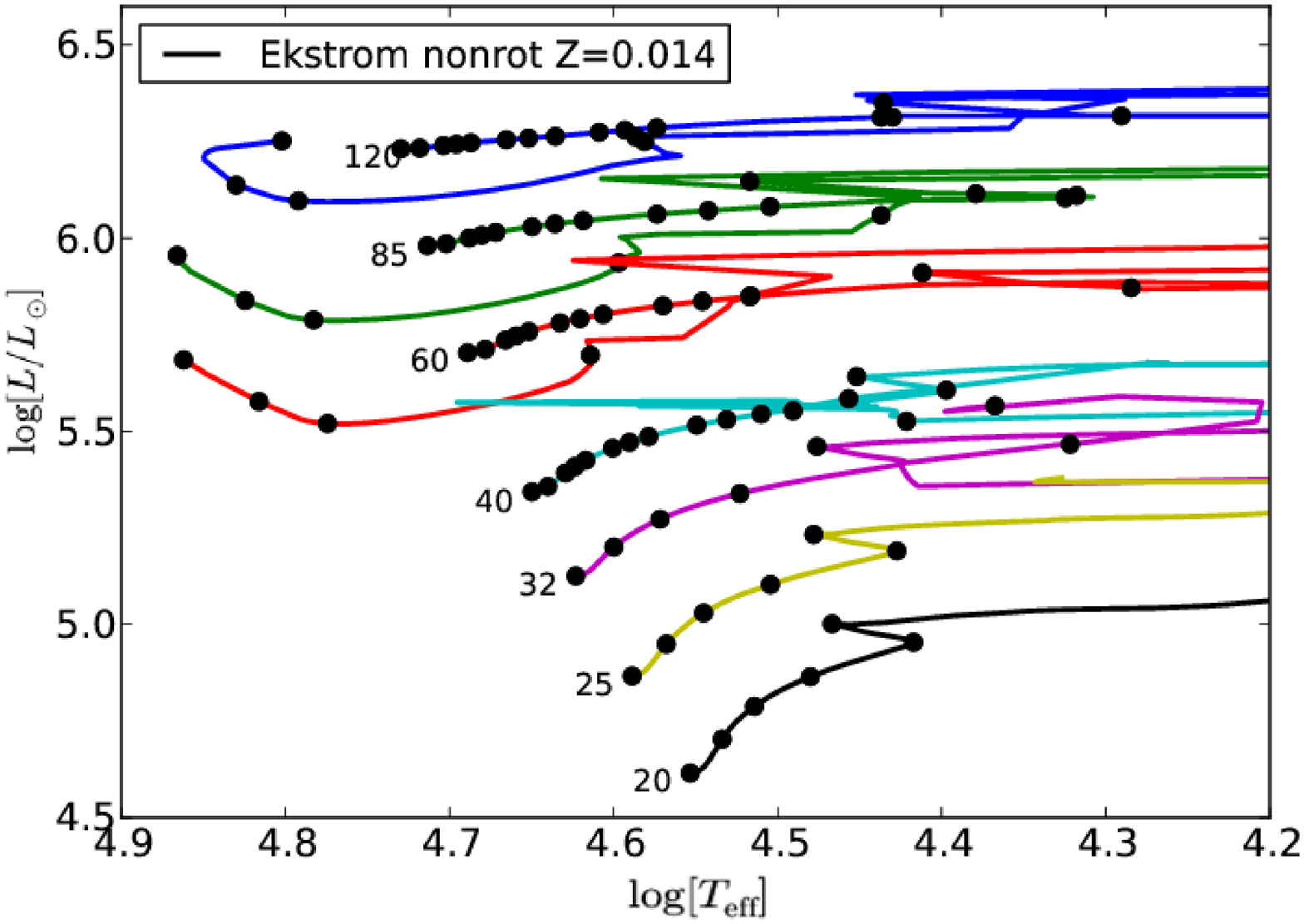}
    \caption{{\it Top}: Evolutionary tracks (Ekstr\"om \etal\ 2012) at solar metallicity computed with rotation and
    labeled by initial mass (in \msun).  
    {\it Bottom}: Corresponding non-rotating tracks.   Points along tracks show values of $T_{\rm eff}$
    and $L$ at which model atmospheres were computed. }
\end{figure}

%%%%%%%%%%%%%%%%%%%%%%%%%%%%%%%%%%%%%%%%%%%%%%%%%%%%%%%%%%

%%%%%%%%%%%%   Figure 6    %%%%%%%%%%%%%%%%%%%%%%%%%%%%%%%%%%%%%%%%%%

\begin{figure}
    \epsscale{1.3} 
    \plotone{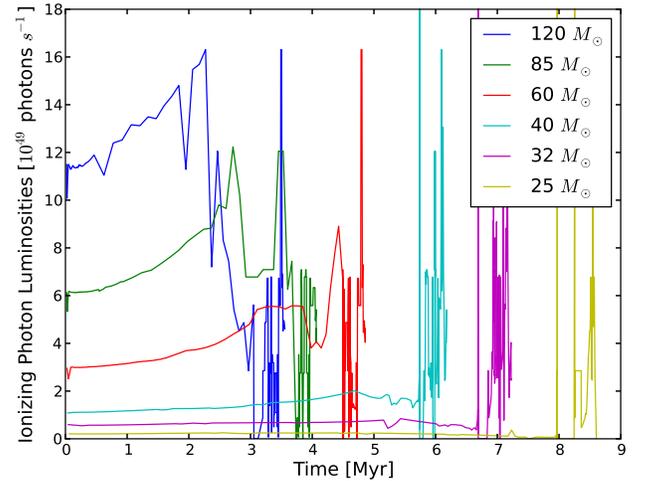}
    \caption{The LyC photon luminosities, $Q_0(t)$, versus time for Ekstr\"om \etal\ (2012) rotating evolutionary tracks
    with $Z = 0.014$.   Each line shows a track with different initial mass from 120~\msun\ to 25~\msun\ (top to bottom).   
    Rapid spikes show ``blue loops" and Wolf-Rayet phase at the end of the star's lifetime.  Although WR stars take up a 
    small fraction of the lifetime, we have ensured that  sufficient model atmospheres are placed to capture the LyC 
    produced in these late phases of evolution. }
\end{figure}

%%%%%%%%%%%%%%%%%%%%%%%%%%%%%%%%%%%%%%%%%%%%%%%%%%%%%%%%%%

Current surveys (Mauerhan \etal\ 2011; Shara \etal\ 2012) identified $\sim 500$ WR stars within 3-4 kpc of the Sun (plus the
Galactic center).  From Galactic distribution models, Shara \etal\ (2009) estimated that the Milky Way could have up to 6500 WR 
stars.  A more recent study (Rosslove \& Crowther 2015) has revised this estimate downward to $1900\pm250$.  Maeder \& 
Meynet (2005) found WR/O $\approx 0.02$ in models with no stellar rotation, increasing to WR/O $\approx 0.05-0.07$ for rotation 
speeds of  300 \kms\ and metallicities $Z = 0.01-0.02$.   The WR/O  ratio drops to only a few percent in the Magellanic Clouds 
(Meynet \& Maeder 2005), although the observed WR/O ratios continue to change with new LMC/SMC surveys (Massey \etal\ 2014).  
Georgy \etal\ (2012) compared recent observations of massive-star populations WR/O $\approx 0.12\pm0.03$ in the Milky Way 
with their solar-metallicity models ($Z = 0.014$) with and without rotation.  
They found that models with rotation could generate WR/O $\approx 0.07$, and they suggested that close-binary 
interactions might explain the remainder.  Thus, we believe that our calculations provide a conservative estimate of 
the LyC produced by O stars and their post-main-sequence successors, consistent with the latest Geneva tracks.  
At the low metallicities expected in the high-redshift galaxies responsible for reionization, the WR populations will likely 
be much smaller. Future calculations may improve the accuracy of the populations by including effects of binaries and 
improved prescriptions for stellar mass loss.   

Our overall goal is to compute the ionizing photon production value at each of the timesteps along the tracks.  Because running 
a model at every time step would require thousands of runs, we chose a different approach.  We ran model atmospheres along 
$\sim100$ key points and then interpolate to define the grid of spectra for each track.  Typically, we have a few tens of models 
for each initial mass, totaling $\sim$100 models for each set of tracks.   For blue loops in two most massive stars (85 $M_{\odot}$ 
and 120 $M_{\odot}$) the stars move far to the left in the H-R diagram, $T_{\rm eff} > 70$~kK. This is too hot for our atmosphere 
code to achieve convergence, as the star exceeds the Eddington luminosity (with full opacity sources).  Instead of interpolating, 
we match the point along the evolutionary track with the closest model atmosphere.  The error of this approximation is small 
because of the short time the stars spend in these regions.  For a general point on a track we choose four model atmospheres 
on the grid surrounding it and linearly interpolate to the track across the parallelogram connecting them in ($T_{\rm eff}$, $\log$) 
space.  Two of the four points are along the same track and provide  the main part of the interpolation.  The two points outside 
the track provide minor corrections.  Some regions of the grid that do not allow us to use all four points for interpolation.  In these 
regions, we use a smaller number of suitable points to complete the process.  There are also times when the evolutionary tracks 
move outside the grid.  We then find the closest grid point and adopt its value for the photon production.  We then integrate the 
ionizing photon luminosities for each initial mass from the evolutionary tracks to find the total number of ionizing photons produced 
over the stellar lifetime.  

\newpage

\section{RESULTS}

\subsection{Lifetime-Integrated \HI\ Photon Luminosities}

Using our coupled tracks and model atmospheres, we derive the total hydrogen LyC photon production, 
$Q_0^{\rm (tot)}(m)$, over the lifetime of individual massive stars (Table~1).    Figure~7 shows these values for three 
{\it older} evolutionary tracks (Schaller \etal\ 1992) with no rotation and at three different metallicities ($Z = 0.004$, 
0.02, 0.04).  These are compared to two {\it new} tracks (Ekstr\"om \etal\ 2012; Georgy \etal\ 2013) at solar metallicity, 
$Z = 0.014$, with and without rotation.  Clearly, the highest LyC production comes from the rotating tracks, which are 
typically 50\% higher, with $Q_0^{\rm (tot)}(m) = (3-11) \times 10^{63}$ photons for initial masses of 40--120~\msun.   
Models calculated at lower metallicity consistently produce more ionizing photons, although the increase is smaller 
($\sim10$\%) than in rotating models.

%%%%%%%%%%%%   Figure 7   %%%%%%%%%%%%%%%%%%%%%%%%%%%%%%%%%%%%%%%%%%

\begin{figure}
    \epsscale{1.1}
    \plotone{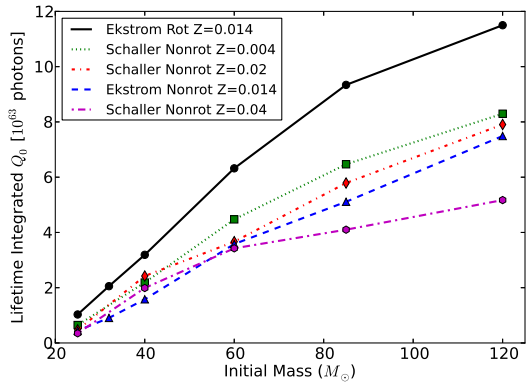}
    \caption{Lifetime-integrated number of H-ionizing photons, $Q_0^{\rm (tot)}(m)$, for three older evolutionary tracks,
    labeled ``Schaerer" and taken from Schaller \etal\ (1992) at three different metallicities.   For comparison, we show 
    LyC production for two new solar-metallicity tracks (Ekstr\"om \etal\ 2012) computed with and without rotation.}
\end{figure}

%%%%%%%%%%%%%%%%%%%%%%%%%%%%%%%%%%%%%%%%%%%%%%%%%%%%%%%%%%%%%%

%%%%%%%%%%%%   Figure 8   %%%%%%%%%%%%%%%%%%%%%%%%%%%%%%%%%%%%%%%%%%

\begin{figure}
   \epsscale{1.1}
    \plotone{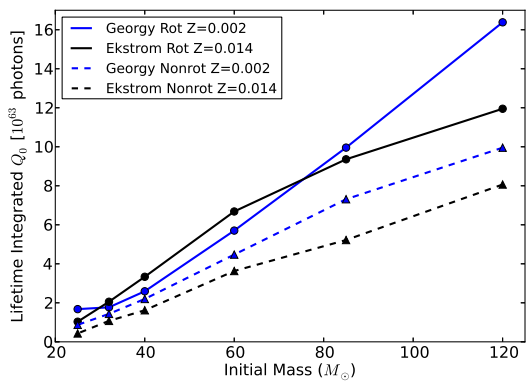}
    \caption{Comparison of lifetime-integrated LyC photon production, $Q_0^{\rm (tot)}(m)$, for new evolutionary tracks 
    (Ekstr\"om \etal\ 2012; Georgy \etal\ 2013), with and without rotation, at solar ($Z = 0.014$) and sub-solar ($Z = 0.002$)
    metallicity.  \vspace{0.4cm} }
\end{figure}

%%%%%%%%%%%%%%%%%%%%%%%%%%%%%%%%%%%%%%%%%%%%%%%%%%%%%%%%%%%%%

Figure~8 compares $Q_0^{\rm (tot)}(m)$ for the {\it new} rotating and non-rotating tracks (Ekstr\"om \etal\ 2012; 
Georgy \etal\ 2013) at the two available metallicities, $Z = 0.014$ and $Z = 0.002$.  Here again, we see that 
LyC production is elevated by $\sim50$\% for models with rotation over most of the mass range of O-stars.  
Our integrations with the {\it new} non-rotating tracks produce $Q_0^{\rm (tot)}(m)$ curves (two dashed lines in 
Figure~8) that are nearly linear with mass, while production rates with the {\it older} tracks (Schaller \etal\ 1992) flatten 
out for the $85\;M_{\odot}$ and $120\;M_{\odot}$ models.  Interestingly, the new rotating-track models of Georgy \etal\ 
(2013) show considerably more LyC photons at $120\;M_{\odot}$, continuing the linear trend of $Q_0^{\rm (tot)}(m)$.
However, those tracks were computed at low metallicity ($Z = 0.002$) which may explain part of the increase in LyC
production.  

In general, stars with lower metallicity have a reduced rate of CNO-cycle burning, which results in a contraction and 
heating of the star's core and increased burning through the proton-proton chain.  Consequently, the surface temperature
is hotter and the star produces more ionizing photons.  Additionally, fewer metals in the atmosphere produce less line 
blanketing, which alters the spectral shape in the UV and EUV.  The effects of stellar rotation on  LyC production and 
$Q_0^{\rm (tot)}(m)$ arise partly from the prolongation of hydrogen burning, as new fuel is mixed into the core.  
Rotating models are more luminous, because of their larger convective cores, and their surfaces are hotter owing to 
enhanced helium abundances.  As our calculations bear out, the ionizing fluxes are greater, as are their lifetime
integrated LyC productions.   Thus, owing to the complicated evolutionary tracks of massive stars through different ranges 
of parameters ($T_{\rm eff}$,  $\log g$, $L$,  $R_*$) there are multiple factors contributing to the increase of LyC production 
found at low metallicity and with enhanced rotation.   

\subsection{Integration Over Initial Mass Functions}

With these data and the IMF of a stellar population, we can calculate the number of LyC photons produced per solar mass 
of star formation.  To do this, we integrate the IMF, $\xi(m)$, multiplied by the lifetime integrated photon number $Q_i(m)$ 
for the three ionization continua ($i=0,1,2$).  We also consider three different IMFs. The first is adapted from Salpeter (1955),
who found $\xi(m) = Km^{-\alpha}$, with index $\alpha = 2.35$ for stars between 0.4~\msun\ and 10~\msun.   We extend 
the mass range to $0.1 \le m \le 120$ ($m$ in \msun) and take $K$ as a normalization constant.  The Salpeter function is
the steepest of the three IMFs, maintaining the same index through its entire mass range, whereas recent surveys suggest 
a turnover at low mass.  The second IMF is from Kroupa (2001), a piecewise-continuous power law, split into three mass 
ranges, each with a different value of $\alpha$.   We integrate over the two stellar mass ranges between 0.08~\msun\ and 
120~\msun, in which $\alpha = 1.3$ ($0.08\le m \le 0.5$) and $\alpha = 2.3$ ($0.5 \le m \le 120$).  The third IMF is from 
Chabrier (2003), with the same index ($\alpha=2.3$) for stars with $m > 0.5$.  For $0.08 < m \leq 0.5$ this IMF has the 
more complicated log-normal form:  $\xi(m)=A\exp[-(\log m-\log m_c)^2/2\sigma^2]$.  

Table~2 summarizes the results of our calculations for various combinations of IMF,  metallicities, and evolutionary tracks,
with and without rotation.   The LyC production efficiency is defined as the lifetime-integrated number of ionizing photons 
per solar mass of star formation.  We list three versions of these parameters, $Q_{\rm LyC}$, $Q_{\rm He I}$ and 
$Q_{\rm He II}$, corresponding to the ionizing continua of \HI, \HeI, and \HeII.  These are sensitive to the sets of model 
atmospheres and stellar tracks, as well as the IMF index ($\alpha$) at the high-mass end, and the integrated mass range 
($m_{\rm min}$ and $m_{\rm max}$).  

\subsection{Reionization and LyC Production Effiicency}

We now apply the new stellar models to reionization by comparing our calculated values of $Q_{\rm LyC}$ to the LyC production 
efficiencies calculated by Shull \etal\ (2012).  The ionizing photons $Q_{\rm LyC}$ produced per \msun\ of star formation are 
important for determining when IGM reionization was mostly complete.  This can be assessed by finding the critical star formation 
rate density (Madau \etal\ 1999)  necessary to keep the IGM ionized, balancing photoionizations with hydrogen recombinations.  
The critical density was recently evaluated (Shull \etal\ 2012) using updated estimates for LyC production rates ($Q_{\rm LyC}$) and
LyC escape fraction ($f_{\rm esc}$) as well as IGM structure (density and temperature).  Hydrogen (case-B) recombination rates 
depend on electron temperature as $T^{-0.845}$ and are enhanced by a ``clumping factor" 
$C_H \equiv  \langle n^2 \rangle / \langle n \rangle^2$.  With calculations of IGM clumping, escape fraction, and 
LyC production efficiencies, Shull \etal\ (2012) suggested fiducial values of $C_H \approx 3$,  $f_{\rm esc} \approx 0.2$,
$T_e = (10^4~{\rm K})T_4$,  and LyC efficicncies $Q_{\rm LyC} = 0.004$ (in $10^{63}$ photons per $M_{\odot}$) to find
\begin{eqnarray}
    \dot{\rho}_{\rm SFR} &=& (0.018\; M_{\odot}\; \rm yr^{-1}\; \rm Mpc^{-3} )\left[\frac{(1+z)}{8}\right]^3  \nonumber \\
         & \times &   \left(\frac{C_{\rm H}/3}{\textit{f}_{\rm esc}/0.2}\right)\left(\frac{0.004}{Q_{\rm LyC}}\right)T^{-0.845}_4   \;   .
\end{eqnarray} 
The value of $f_{\rm esc}$ for low-mass galaxies with a cloudy ISM has been estimated to be 10-20\% (Fernandez \& Shull 2011).
Depending on choices of $C_H$ and $T_e$, the escape fraction needs to be at least 10\% in order to complete IGM 
reionization by $z=7$ (Shull \etal\ 2012; Robertson \etal\ 2013; Finkelstein \etal\ 2013).  

In our study, we have explored three factors that affect the LyC production efficiency:  (1) decreases in stellar metallicity; 
(2) increases in stellar rotation;  (3) shape of the IMF. 
From our new calculations of ionizing photon production rates we assess how LyC production efficiency is affected by
choices of stellar tracks (rotation, metallicity), model atmospheres, and IMF.  By comparing results for various combinations
of these parameters, we findthe general trends with metallicity and rotation.   We currently have the full range of evolutionary 
tracks at {\it two} metallicities, solar and sub-solar.  We integrate over three IMFs, labeled in Table~2 as $(1, 2, 3)$ corresponding 
to (Salpeter, Kroupa, Chabrier).   The stellar tracks are labeled as $(4, 5 ,6)$  corresponding to (Schaller 1992, Ekstr\"om 2012, 
Georgy 2013).  The latter two tracks come with and without rotation.   Our results (Figures 7 and 8) show that for each set of 
non-rotating evolutionary tracks their rotating counterparts produce more photons.  Metallicity is less important for the overall 
efficiency, $Q_{\rm LyC}$, although it can be a factor at specific masses.   For the 40 \msun\ and 60 \msun\ models 
with rotation, the solar metallicity models produce more photons than the sub-solar models.  Owing to the power-law 
decrease of the IMF at high masses, this small difference in ionizing photon production results in the solar-metallicity models 
producing slightly more photons per solar mass.   For fixed evolutionary tracks, metallicities, and model atmospheres, adopting  
the Kroupa or Chabrier IMF increases the LyC efficiency by factors of 1.56 and 1.64, respectively, relative to the Salpeter IMF.  

We now perform a detailed comparison of our computed values of $Q_{\rm LyC}$ to the fiducial value $Q_{\rm LyC}=0.004$ 
suggested by Shull \etal\ (2012).   With different combinations of IMFs and evolutionary tracks, $Q_{\rm LyC}$ extends from 
0.00315 to 0.00940 for solar metallicity and from 0.00402 to 0.00910 for sub-solar metallicity (Table~2).  This spread in efficiencies
translates to a factor of 3 in critical SFR density.  
Values of $Q_{\rm LyC}$ computed with rotating evolutionary tracks and the Kroupa IMF were at the upper end of the range.   
One can further analyze these $Q$-values to separate the effects of reduced metallicity from those of rotation.  The three IMFs only
change the overall normalization, because of their different mass-loading by low-mass stars.  Because the Salpeter IMF includes 
more low-mass stars, the absolute values of $Q_{\rm LyC}$ are reduced;  these stars add mass but produce few ionizing photons.   

First, we examine the effects of metallicity.    For the ``old"  non-rotating tracks (Schaller \etal\ 1992) at solar metallicity ($Z = 0.020$) 
we find $Q_{\rm LyC} = 0.00367$, 0.00572, and 0.00370 for the three IMFs (1,2,3) considered here (Salpeter, Kroupa, and Chabrier,
respectively).  For the new  non-rotating tracks (Ekstr\"om \etal\ 2012) at sub-solar metallicity ($Z = 0.004$) we find 
$Q_{\rm LyC} = 0.00399$, 0.00623, and 0.00402 for IMFs (1,2,3).  Taking the appropriate ratios for the old tracks and corresponding 
model atmospheres,  we find that decreasing the metallicity from solar ($Z = 0.02$) to sub-solar ($Z = 0.004$) increases the LyC 
efficiencies by only 9\%.   That is, the ratio $Q_{\rm LyC}({\rm sub solar}) / Q_{\rm LyC}({\rm solar}) = 1.09$ for all three IMFs.   However, 
performing the same calculation for the new non-rotating tracks (Ekstr\"om \etal\ 2012; Georgy \etal\ 2013), we find a 34\% increase 
at sub-solar metallicity, with $Q_{\rm LyC}({\rm sub solar}) / Q_{\rm LyC}({\rm solar}) = 1.34$.   For the new rotating tracks, the LyC 
efficiencies at solar and sub-solar metallicity are essentially the same, with
$Q_{\rm LyC}({\rm sub solar}) / Q_{\rm LyC}({\rm solar}) = 0.97$ for all three IMFs.  
Chabrier and Kroupa IMFs have the same power-law slope for high mass ($\alpha=2.3$), but each uses a different fit for low-mass 
stars.   The Chabrier IMF has more low-mass stars compared to the Kroupa IMF and therefore more stellar mass that is not producing 
ionizing photons.  
Next, we examine the effect of rotation on $Q_{\rm LyC}$ at fixed metallicity.  For all three IMFs,  the evolutionary tracks with 
rotation produce more ionizing photons per solar mass.  At solar metallicity ($Z = 0.014$) the Ekstr\"om \etal\ (2012) tracks
and corresponding model atmospheres yield 87\% higher LyC efficiency, with
$Q_{\rm LyC}({\rm rot}) / Q_{\rm LyC}({\rm non rot}) = 1.87$.  At sub-solar metallicity ($Z = 0.002$) the Georgy \etal\ (2013) tracks 
and corresponding atmospheres were 35\% more efficient in LyC production, with 
$Q_{\rm LyC}({\rm rot}) / Q_{\rm LyC}({\rm nonrot}) = 1.35$.  

As a general conclusion, we find that increased stellar rotation is more important than reduced metallicity in boosting the LyC  
efficiency.  By comparing subsets of tracks, metallicities, and IMFs, we can recover more specific results.  We label the LyC 
production efficiencies by $Q_{\rm LyC}{\rm (IMF, Tracks)}$, where IMFs and tracks are defined as in Table~2 (1 = Salpeter,  
2 = Kroupa,  3 = Chabrier) with non-rotating tracks labeled by (4 = Schaller; 5 = Ekstr\"om) and rotating tracks by
(5 = Ekstr\"om, 6 = Georgy).  For the relevant combinations of parameters, we find the following efficiencies for  
``non-rotating, solar metallicity" stellar populations:  
\begin{eqnarray}
      Q_{\rm LyC}(1,4) = 0.00367 \; \; Q_{\rm LyC}(2,4) = 0.00572    \nonumber \\  
        Q_{\rm LyC}(3,4) = 0.00370   \; \; (Z = 0.020) \nonumber  \\
      Q_{\rm LyC}(1,5) = 0.00322 \; \; Q_{\rm LyC}(2,5) = 0.00503      \nonumber \\  
       Q_{\rm LyC}(3,5) = 0.00325   \; \; (Z = 0.014)   
 \end{eqnarray}
 The range of $Q_{\rm LyC}$ between $0.00322 - 0.00572$ arises primarily from the choice of IMF, which affects the 
 ``mass-loading" at the low end.  
 Next, we make a similar tabulation for ``non-rotating, sub-solar metallicity" populations (either $Z = 0.004$ or $0.002$):  
\begin{eqnarray}
      Q_{\rm LyC}(1,4) = 0.00399  \; \;  Q_{\rm LyC}(2,4) = 0.00623     \nonumber \\
      Q_{\rm LyC}(3,4) = 0.00402    \; \; (Z = 0.004)  \nonumber  \\
      Q_{\rm LyC}(1,6) = 0.00433  \; \;  Q_{\rm LyC}(2,6) = 0.00676    \nonumber \\
       Q_{\rm LyC}(3,6) = 0.00436    \; \; (Z = 0.002)  
 \end{eqnarray} 
As before, the range of $Q_{\rm LyC}$ from $0.00399 - 0.00676$  arises from the assumed IMF.  
 Finally,  for the rotating tracks at both solar and sub-solar metallicities, we find:  
 \begin{eqnarray}
      Q_{\rm LyC}(1,5) = 0.00603  \; \; Q_{\rm LyC}(2,5) = 0.00940   \nonumber \\
      Q_{\rm LyC}(3,5) = 0.00607    \; \; (Z = 0.014)  \nonumber  \\
      Q_{\rm LyC}(1,6) = 0.00583  \; \; Q_{\rm LyC}(2,6) = 0.00910  \nonumber \\
      Q_{\rm LyC}(3,6) = 0.00588    \; \; (Z = 0.002)  
 \end{eqnarray} 
 The range of $Q_{\rm LyC}$ from $0.00583 - 0.00940$ arises mainly from the assumed IMF.  

Adopting an IMF with fewer low-mass stars or using tracks with increased stellar rotation produces larger boosts in LyC efficiency 
than sub-solar metallicity.  For non-rotating stars at solar metallicity, a typical value is $Q_{\rm LyC} =  0.004\pm 0.002$, as found
in our previous study (Shull \etal\ 2012).   Including stellar rotation could increase the efficiency to 
$Q_{\rm LyC} =  0.0075\pm 0.0015$, although those stellar tracks assumed a single value of rotation at 40\% of breakup, which 
may be excessively high.  Allowing for the range in $Q_{\rm LyC}$ efficiencies produced by assumptions about IMF, metallicity, 
and rotation, we recommend adopting an overall LyC production efficiency of $Q_{\rm LyC} =  0.006\pm 0.002$ in Equation (3), 
or a photon production calibration of $(6\pm2) \times 10^{60}$ LyC photons per \msun\ of star formation.  
This is equivalent to a SFR calibration of $10^{53.3\pm0.2}$ photons~s$^{-1}$ for SFR in \msun~yr$^{-1}$ and represents  a 
50\% increase over previous estimates.  With the new LyC rates, the critical SFR to maintain reionization can be rewritten
\begin{eqnarray}
    \dot{\rho}_{\rm SFR} &=& (0.012\; M_{\odot}\; \rm yr^{-1}\; \rm Mpc^{-3})\left[\frac{(1+z)}{8}\right]^3 \nonumber \\
        & \times &   \left(\frac{C_{\rm H}/3}{\textit{f}_{\rm esc}/0.2}\right)\left(\frac{0.006}{Q_{\rm LyC}}\right)T^{-0.845}_4   \;   .
\end{eqnarray} 
The 50\% boost in LyC production efficiency is an important ingredient in assessing whether IGM reionization was complete
by $z \approx 7$, as suggested by recent observations (Oesch \etal\ 2014; McLure \etal\ 2013; Robertson \etal\ 2013; 
Finkelstein \etal\  2012).

%%%%%%%%%%%%   Figure 9a ,b    %%%%%%%%%%%%%%%%%%%%%%%%%%%%%%%%%%%%%%%%%%

\begin{figure} 
   \epsscale{1.1}
    \plotone{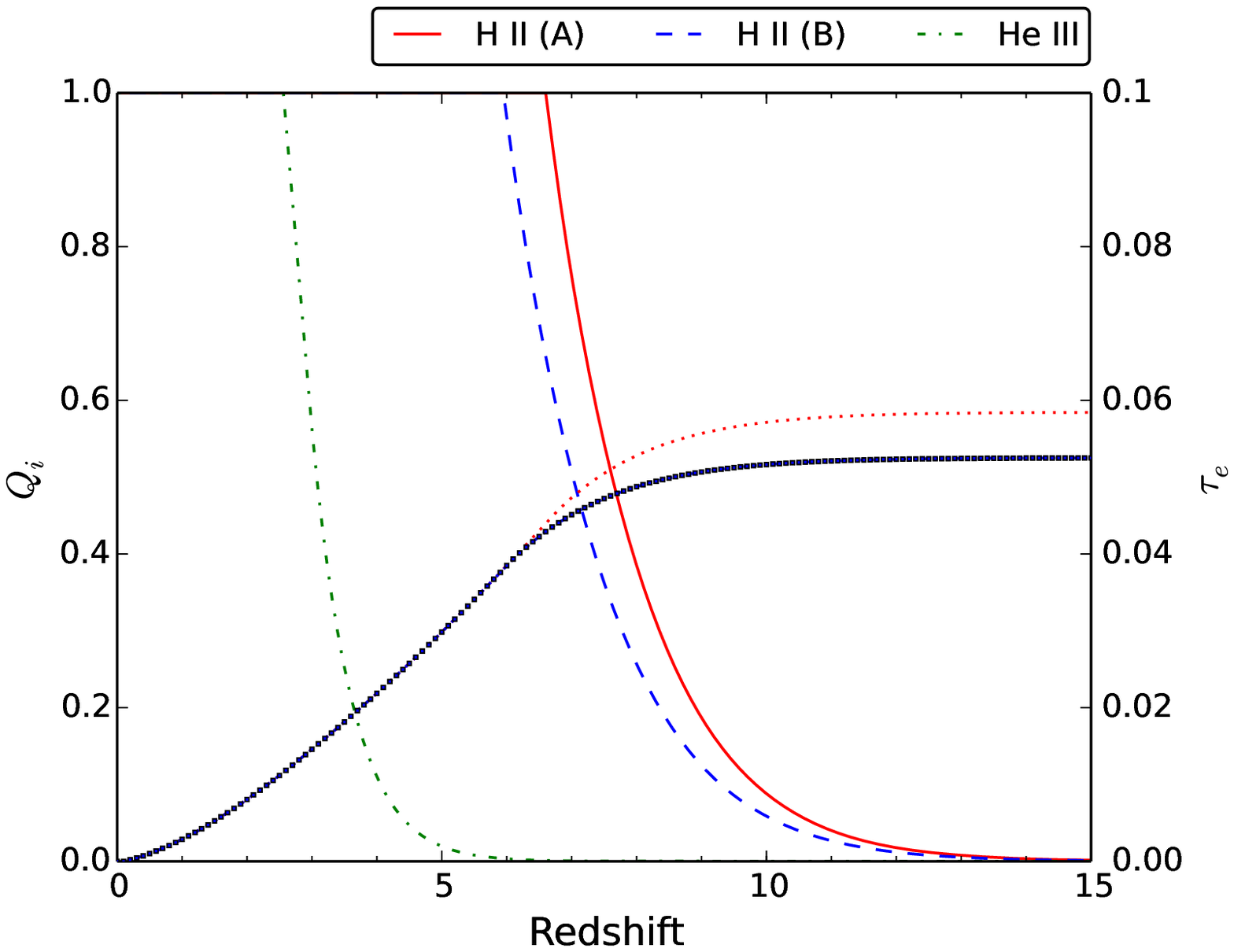}
  \epsscale{1.0}
 \plotone{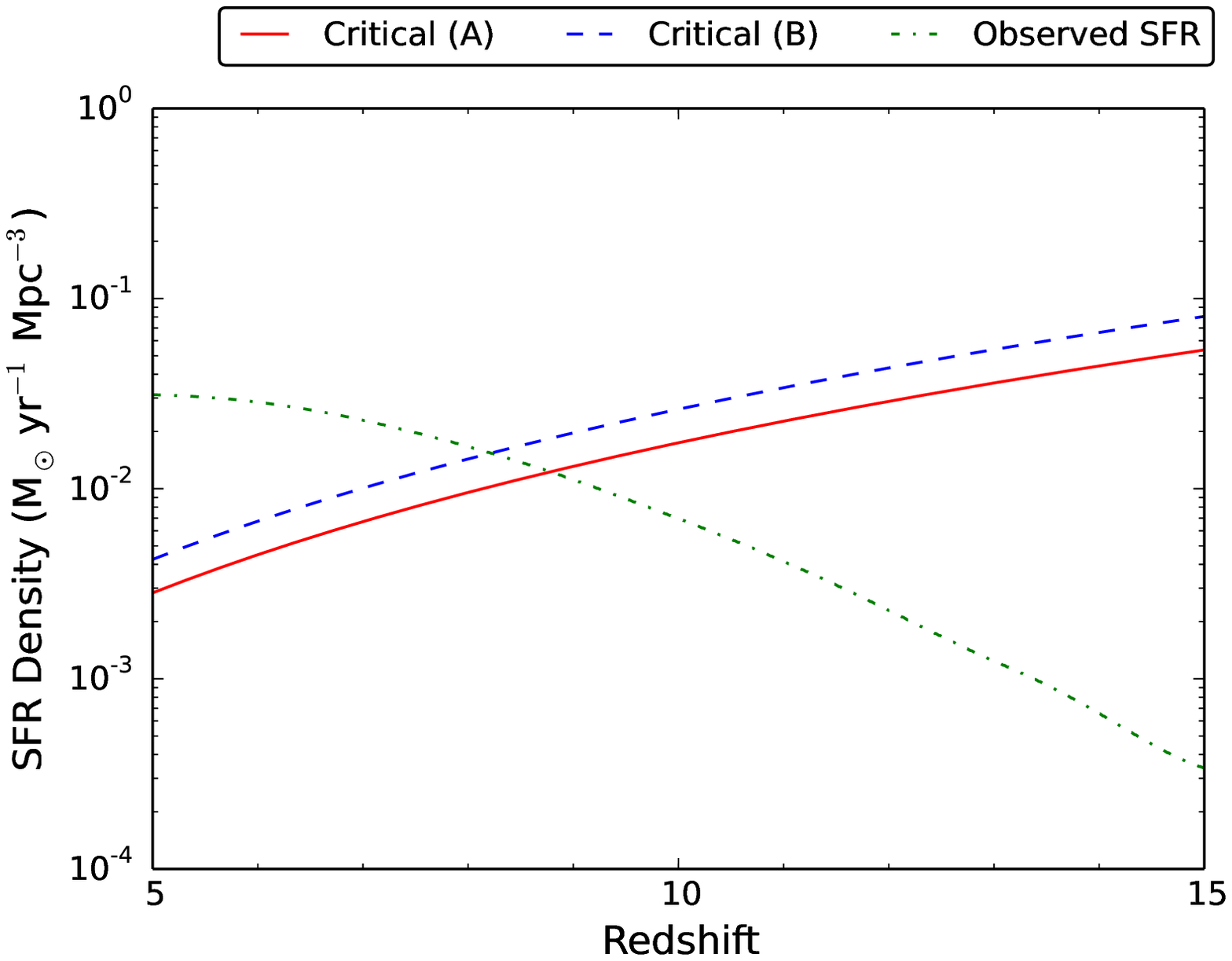}
    \caption{Reionization histories  and critical SFRs  for two LyC production efficiencies, computed with our 
     reionization simulator (Shull \etal\ 2012).   {\it Left:}  Ionized volume fraction $Q_i$ and integrated CMB optical 
     depth, $\tau_e$, to electron scattering for two LyC production rates:  $Q_{\rm LyC} = 0.006$ (solid red curve, model~A) 
     and $Q_{\rm LyC} = 0.004$ (dashed blue curve, model~B), both in $10^{63}$ photons per \msun\ of star formation. 
     The dot-dashed curve at $z < 5$ follows the \HeII\ history, governed by quasars and completed at $z \approx 3$.  
     {\it Right:}  Critical SFR densities required to maintain IGM ionization compared to observed and extrapolated SFR 
     (green dot-dashed curve).  Reionization models (assume clumping factor $C_H = 3$, LyC escape 
     fraction $f_{\rm esc} = 0.2$, and electron temperature $T_e = 20,000$~K.  The SFR history comes from the galaxy
     luminosity  function (Trenti \etal\ 2010) extrapolated down to absolute magnitude $M_{\rm AB} = -10$ in rest-frame 
     UV and converted to SFR density using the prescription in Madau \etal\ (1999).    \vspace{0.6cm}
         }
\end{figure}

%%%%%%%%%%%%%%%%%%%%%%%%%%%%%%%%%%%%%%%%%%%%%%%%%%%%%%%%%%%%%%

These considerations are illustrated in Figure 9, which shows the reionization histories and critical SFR densities derived
with the reionization simulator developed in a previous study (Shull \etal\ 2012).  The left panel shows the redshift
evolution of the ionized volume fraction, $Q_i(z)$, for two values of LyC production efficiency, $Q_{\rm LyC} = 0.004$ (the old
standard) and $Q_{\rm LyC} = 0.006$ (the new efficiency).  Accompanying curves show the evolution of CMB optical depth, 
$\tau_e$, to electron scattering, an integral constraint on the ionization history.  These two models, labeled A and B, adopt 
identical physical parameters, $C_H = 3$, $f_{\rm esc} = 0.2$, and $T_e = 20,000$~K.  The SFR history follows that of
Trenti \etal\ (2010) with the galaxy luminosity function extrapolated down to absolute magnitude $M_{\rm AB} = -10$
in the rest-frame UV.   The curve at $z < 5$ follows the \HeII\ history, which is governed by quasars, with \HeII\ reionization
completed at $z \approx 3$.  The right panel compares $\dot{\rho}_{\rm SFR}$, the critical SFR density from Equation (7), to 
the observed (and extrapolated) SFR density.  The increased LyC production efficiency shifts the evolution in ionized volume 
fraction, $Q_i(z)$, toward higher redshifts.  One can therefore offset $Q_{\rm LyC} \approx 0.006$ against the combination 
of physical parameters, $(C_H / f_{\rm esc}) T_4^{-0.845} \approx 8.35$ in this case.   With the elevated LyC efficiency 
(model~A), full reionization ($Q_i \approx 1$) occurs at $z \approx 7$, and the  ``observed and extrapolated SFR exceeds 
$\dot{\rho}_{\rm SFR}$ at $z < 9$.   The ionization history can be altered with various combinations of physical parameters 
and by different extrapolations to low-luminosity galaxies.

\subsection{\HeI\ and \HeII\ Continuum Photon Luminosities}

From the same data used to determine LyC photon fluxes, we only need to change integration limits to compute fluxes in the 
\HeI\ and \HeII\ continua. The calculation of $Q_1(m)$, the \HeI-ionizing photon production for each initial mass, follows the same 
procedure as for $Q_0(m)$ using Equation (1) with $\lambda_{\rm lim} = 504.259$ \AA.  The $Q_1(m)$ curves with solar metallicity 
(Figure~10) have a similar shape to the \HI\ Lyman continuum curves in Figures~7 and 8.  In both cases, the photon production increases 
with mass but flattens out at $120\;M_{\odot}$.  Because the photons are in the Wien limit ($h\nu \gg kT_{\rm eff}$) of the 
Planck function, the intensity of radiation decreases exponentially in this short-wavelength range. Consequently the photon 
production rates for \HeI\ ionization are significantly less than the \HI\ continuum rates, by up to an order of magnitude.
The comparison between the new evolutionary tracks at two different metallicities follows a different trend.  The first difference is 
the behavior of the two most massive models in the sub-solar rotating model.  Instead of flattening with increasing mass, the models 
seem to produce a rapidly increasing number of ionizing photons.  The other sets of evolutionary tracks produce rather linear 
$Q_1(m)$ curves.

%%%%%%%%%%%%   Figure 10   %%%%%%%%%%%%%%%%%%%%%%%%%%%%%%%%%%%%%%%%%%

\begin{figure}
    \epsscale{1.2}
    \plotone{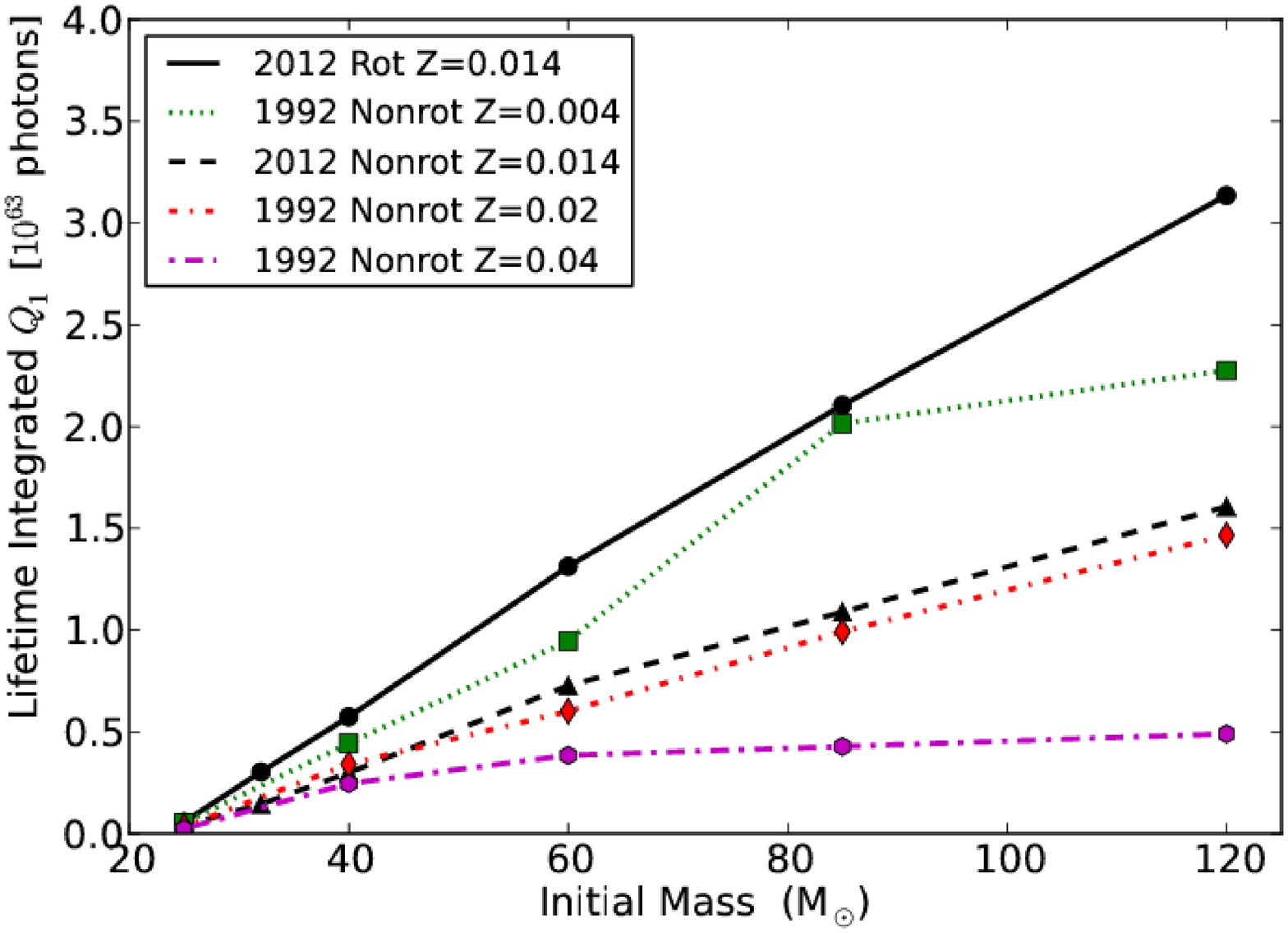}
    \plotone{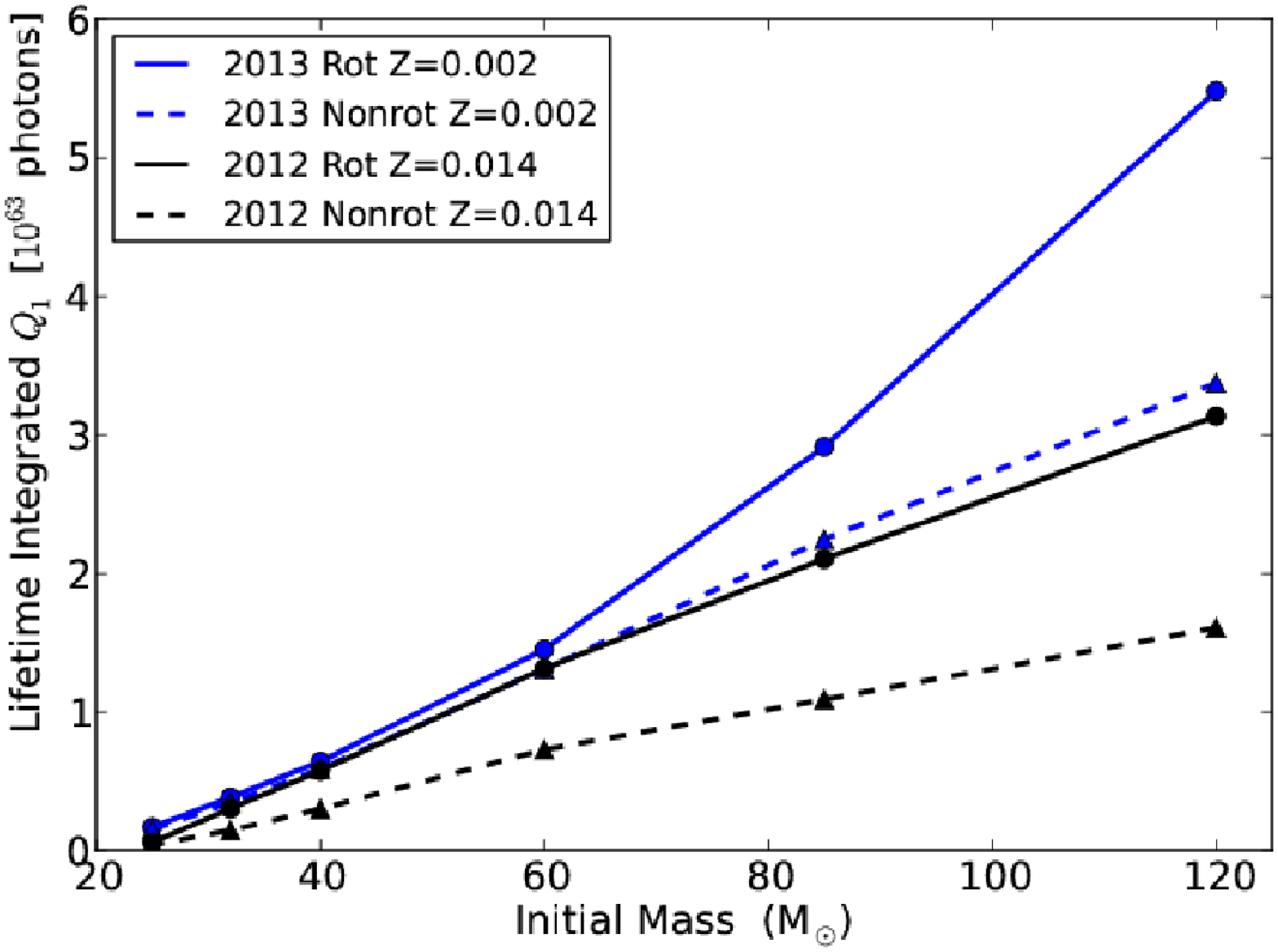}
    \caption{Lifetime-integrated \HeI\ continuum photon production $Q_1^{\rm (tot)}(m)$ in $10^{63}$ photons vs.\ initial mass 
    for several evolutionary tracks.  {\it Top}:  former standard tracks compared to new rotating and non-rotating solar metallicity 
    tracks.  ``1992 Nonrot" tracks from Schaller \etal\ (1992), and ``2012 Rot/Nonrot" tracks from Ekstr\"om \etal\ (2012).  
    {\it Bottom}:  new rotating and non-rotating tracks at solar and sub-solar metallicity.   ``2013 Rot/Nonrot" tracks are 
    from Georgy \etal\ (2013). \vspace{0.4cm} }

\end{figure}

%%%%%%%%%%%%%%%%%%%%%%%%%%%%%%%%%%%%%%%%%%%%%%%%%%%%%%%%%%%%%%%

%%%%%%%%%%%%   Figure 11  %%%%%%%%%%%%%%%%%%%%%%%%%%%%%%%%%%%%%%%%%%

\begin{figure}
    \epsscale{1.1}
    \plotone{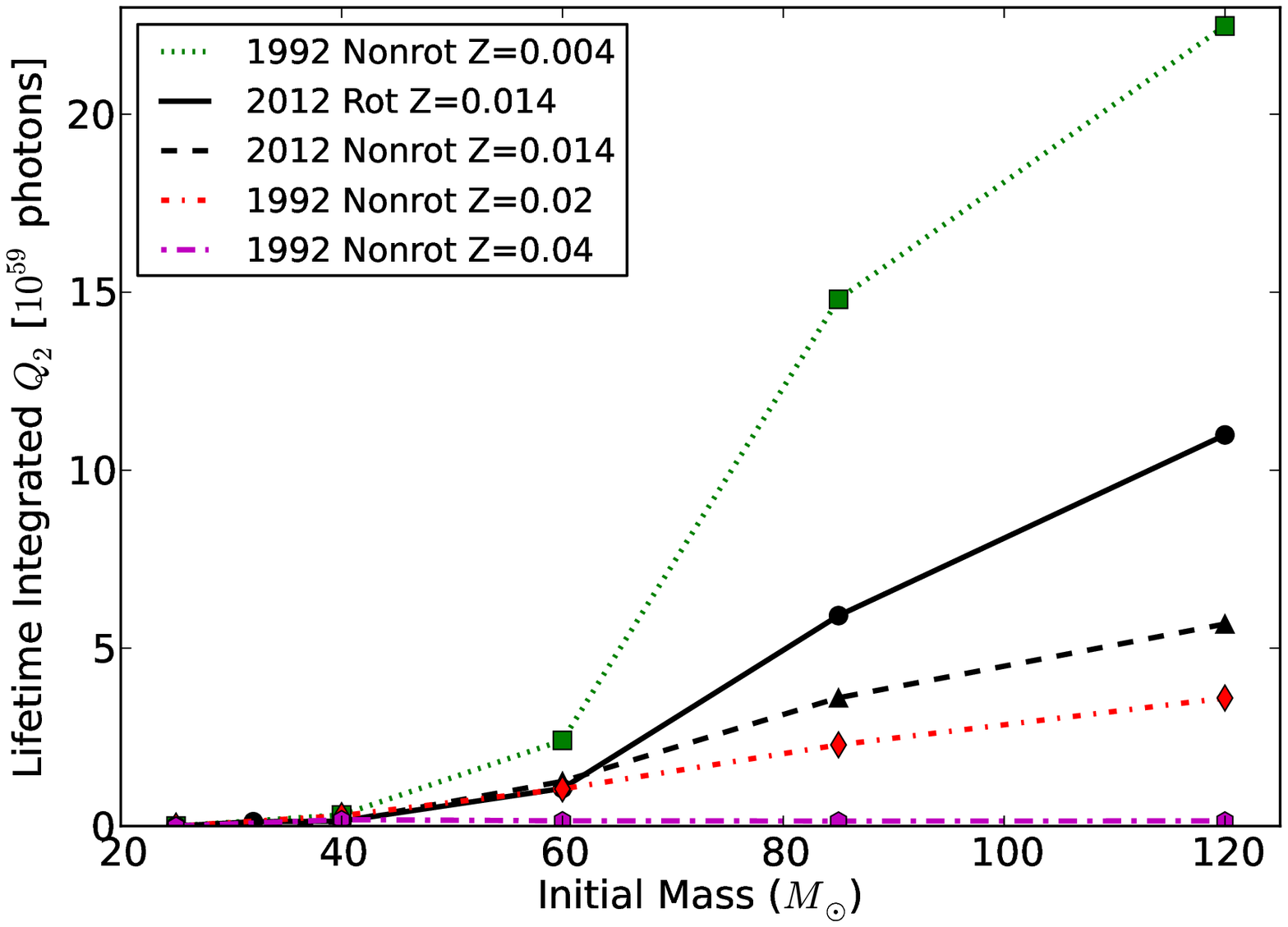}
    \plotone{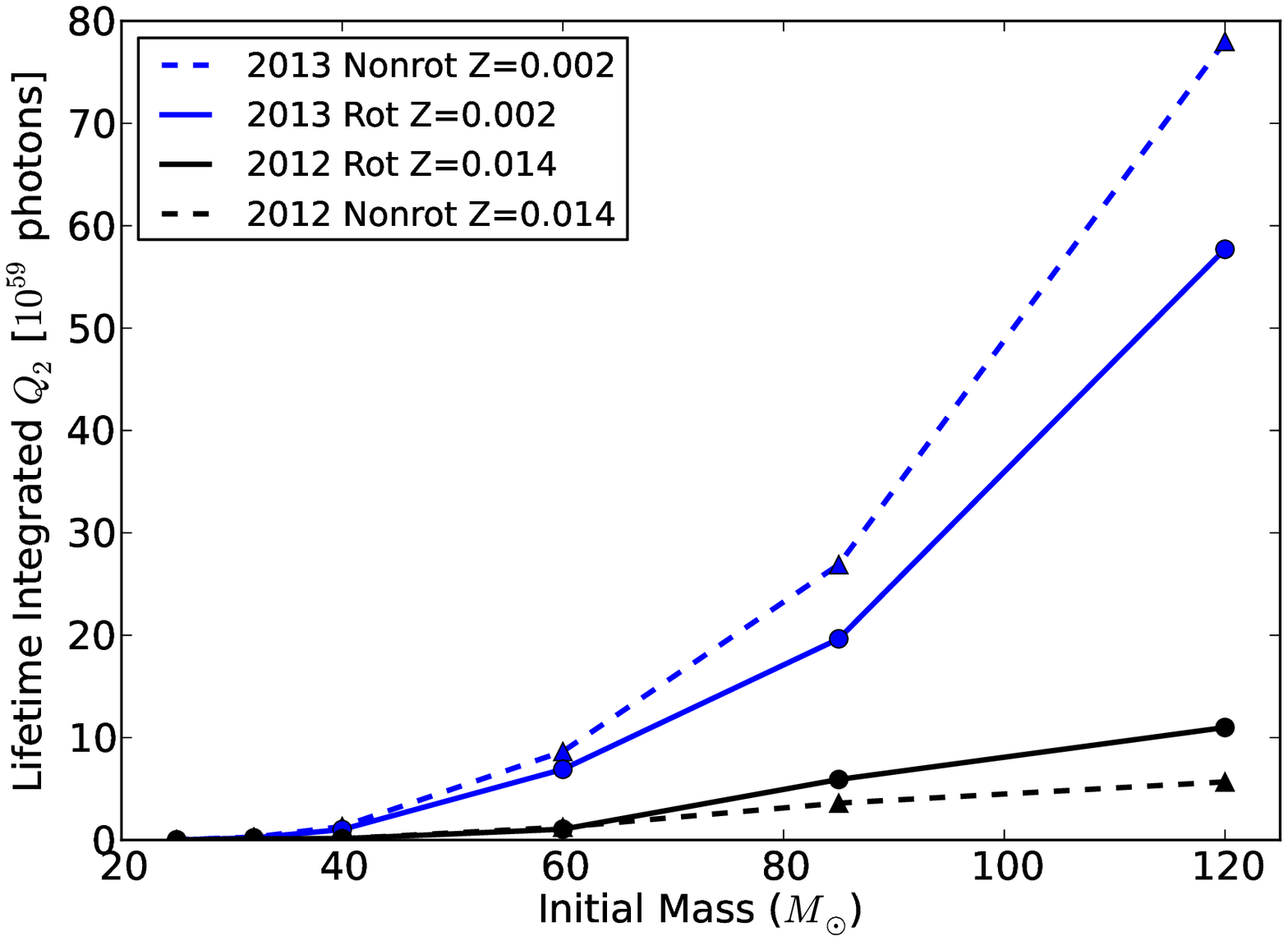}
    \caption{Lifetime-integrated \HeII\ continuum photon production $Q_2^{\rm (tot)}(m)$ vs.\ initial mass.  
    {\it Top panel}:  Comparison of old evolutionary tracks (Schaller \etal\ 1992) and new solar metallicity rotating and 
    non-rotating evolutionary tracks (Ekstr\"om \etal\ 2012).
    {\it Bottom}:  Comparison of new rotating and non-rotating evolutionary tracks at solar and sub-solar metallicity 
    (Ekstr\"om \etal\ 2012; Georgy \etal\ 2013).  \vspace{0.4cm} }
\end{figure}    

%%%%%%%%%%%%%%%%%%%%%%%%%%%%%%%%%%%%%%%%%%%%%%%%%%%%%%%%%%%%%

From the data in Table~2, we find a factor of three range in $Q_{\rm HeI}$ depending on choice of IMF, metallicity, and rotation.  For the 
``old" non-rotating models (Schaller \etal\ 1992 tracks) we find $Q_{\rm HeI} \approx (0.5-0.9) \times 10^{60}$ photons per \msun\ at 
solar metallicity.  At sub-solar metallicities, the \HeI\ continuum production is higher by 60\% with 
$Q_{\rm HeI}({\rm sub solar}) / Q_{\rm HeI}({\rm solar}) = 1.6$.  In the newer non-rotating tracks (Ekstr\"om \etal\ 2012 and
Georgy \etal\ 2013) $Q_{\rm HeI}$ is a factor of 2 larger.  For the new rotating models, the range is
$Q_{\rm HeI} \approx (1.1-1.7) \times 10^{60}$ photons per \msun\ at solar metallicity and 32\% higher at sub-solar metallcity.  
Rotation elevates  $Q_{\rm HeI}$ at fixed metallicity for all three IMFs.   At solar metallicity ($Z = 0.014$) the 
Ekstr\"om \etal\ (2012) tracks and corresponding atmospheres yield 94\% higher efficiency, with
$Q_{\rm HeI}({\rm rot}) / Q_{\rm HeI}({\rm non rot}) = 1.94$.  At sub-solar metallicity ($Z = 0.002$) the Georgy \etal\ (2013) tracks 
and corresponding atmospheres were 22\% more efficient, with $Q_{\rm HeI}({\rm rot}) / Q_{\rm HeI}({\rm nonrot}) = 1.22$.  
The overall \HeI\ efficiency ranges are comparable to those for $Q_{\rm LyC}$.  The recommended median value is
$Q_{\rm HeI} = 0.6 \times 10^{60}$ photons per \msun\ (non-rotating, solar metallicity) with significant enhancements (60-110\%)
at lower metallicity, unlike for the total LyC.  The enhancements in $Q_{\rm HeI}$ due to rotation are comparable to those for
the LyC, at both solar metallicity (90\% boosts) and sub-solar (22-35\% boosts).  

Finally, we compute the  \HeII-ionizing photon fluxes $Q_2(m)$ using Equation (1) with $\lambda_{\rm lim} = 227.838$ \AA.  At
higher ionization energies, the production of \HeII-continuum photons is several orders of magnitude lower than for \HI\ or \HeI\ 
continua.  Curves of $Q_2(m)$ vs.\ initial mass are shown in Figure~11 for rotating and non-rotating evolutionary tracks at solar 
and sub-solar metallicity.   Several features distinguish $Q_2(m)$ from $Q_0(m)$ or $Q_1(m)$ curves.  First, the lower metallicity 
models produce significantly more photons than models with higher metallicity.  This is consistent with our expectations and the
results of the $Q_0(m)$ and $Q_1(m)$ calculations.  The next distinction is that at sub-solar metallicity, the non-rotating models 
produce more \HeII\ continuum photons than corresponding rotating models.  This is different from our expectations and the trends 
from solar-metallicity models.  We can understand this discrepancy by examining the behavior of the sub-solar metallicity tracks, 
in which non-rotating models have a higher effective temperature and luminosity for roughly 1.5 Myr, which constitutes a majority of 
their lifetimes.  At the high energies of the \HeII\ continuum, the spectrum is quite sensitive to $T_{\rm eff}$.  

The recommended value for \HeII\ is $Q_{\rm HeII} = 1 \times 10^{56}$ photons per \msun\ (non-rotating, solar metallicity).
The enhancements in $Q_{\rm HeII}$ due to rotation are a factor of 1.48 at solar metallicity, but reduced by a factor of 0.76 at 
sub-solar metallicity.  Using the same pairwise comparisons as used above for $Q_{\rm LyC}$ and $Q_{\rm HeI}$, we find
large boosts in $Q_{\rm HeII}$ at lower metallicity. The ratio $Q_{\rm HeII}({\rm sub solar}) / Q_{\rm HeII}({\rm solar}) = 4.65$ 
for old non-rotating tracks (and corresponding atmosphere models) as well as for the new rotating tracks.   \\

\subsection{LyC Spectra in Model OB Associations}

The spectral evolution of an ensemble of stars of various masses can be combined through an IMF and followed over time.  
Such spectral synthesis codes are valuable tools for ultraviolet spectral libraries (Robert \etal\ 1993; Rix \etal\ 2004; Leitherer 
\etal\ 2014) and stellar population synthesis codes such as \texttt{Starburst99} (Leitherer \etal\ 1999).  Figure 12 shows a
series of spectra for a cluster starburst containing $10^5~M_{\odot}$ in which the stars follow a Salpeter IMF
($0.1 \leq m \leq 120$).  The two panels show model spectra at times $t =$ 0, 1, 3, 5, and 7 Myr after a coeval burst of star 
formation, using evolutionary tracks with rotation, at solar metallicity ($Z = 0.014$ from Ekstr\"om \etal\ 2012) and at sub-solar 
metallicity ($Z = 0.014$ from Georgy \etal\ 2013).   These models were created by Monte-Carlo sampling of stars for each mass 
for which we have an evolutionary track ($m = 20$, 25, 32, 40, 60, 85, 120).  For the Salpeter differential mass distribution, 
$\xi(m) = Km^{-\alpha}$ with $\alpha = 2.35$, the fraction of stars above mass $m$ is given by
\begin{equation}
   \frac {N(>m)} {N} = \frac { [ m^{-(\alpha-1)} - m_{\rm max}^{-(\alpha-1)} ] }  
                                             { [ m_{\rm min}^{-(\alpha-1)} - m_{\rm max}^{-(\alpha-1)} ] }  \; \; .
\end{equation} 
The constant $K$ is normalized to the total cluster mass $M = 5.862K$, and the total number of stars $N = 16.582K$ for 
mass limits $m_{\rm min} = 0.1$ and $m_{\rm max} = 120$.  For $M = 10^5~M_{\odot}$, we expect a mean number 
$\bar{N} = 282,500$ stars and mean stellar mass $\langle m \rangle = M/N = 0.354$~\msun.  In many of our Monte-Carlo 
samples, the numbers fluctuate about this value, with small numbers at the high-mass end of the IMF.  The mean numbers
of stars at the high-mass end are:  $N(>60~M_{\odot}) \approx 30$ and $N(>100~M_{\odot}) \approx 5$.    
If the upper mass was extended to $m_{\rm max} = 200$, we would expect 40 stars above 60 \msun\ and 15 stars above 
100~\msun.    As discussed earlier, the existence of these very massive stars is controversial, owing to resolution effects.
Moreover, the LyC from stars at $m > 100$~\msun\ may not escape the embedded cloud from which they formed, or
from the dense gas produced in mass-loss episodes or binary mergers (Smith 2014).   We therefore do not consider
stars above the 120~\msun\ track.

%%%%%%%%%%%%   Figure 12   %%%%%%%%%%%%%%%%%%%%%%%%%%%%%%%%%%%%%%%%%%

\begin{figure}
    \epsscale{1.2}
    \plotone{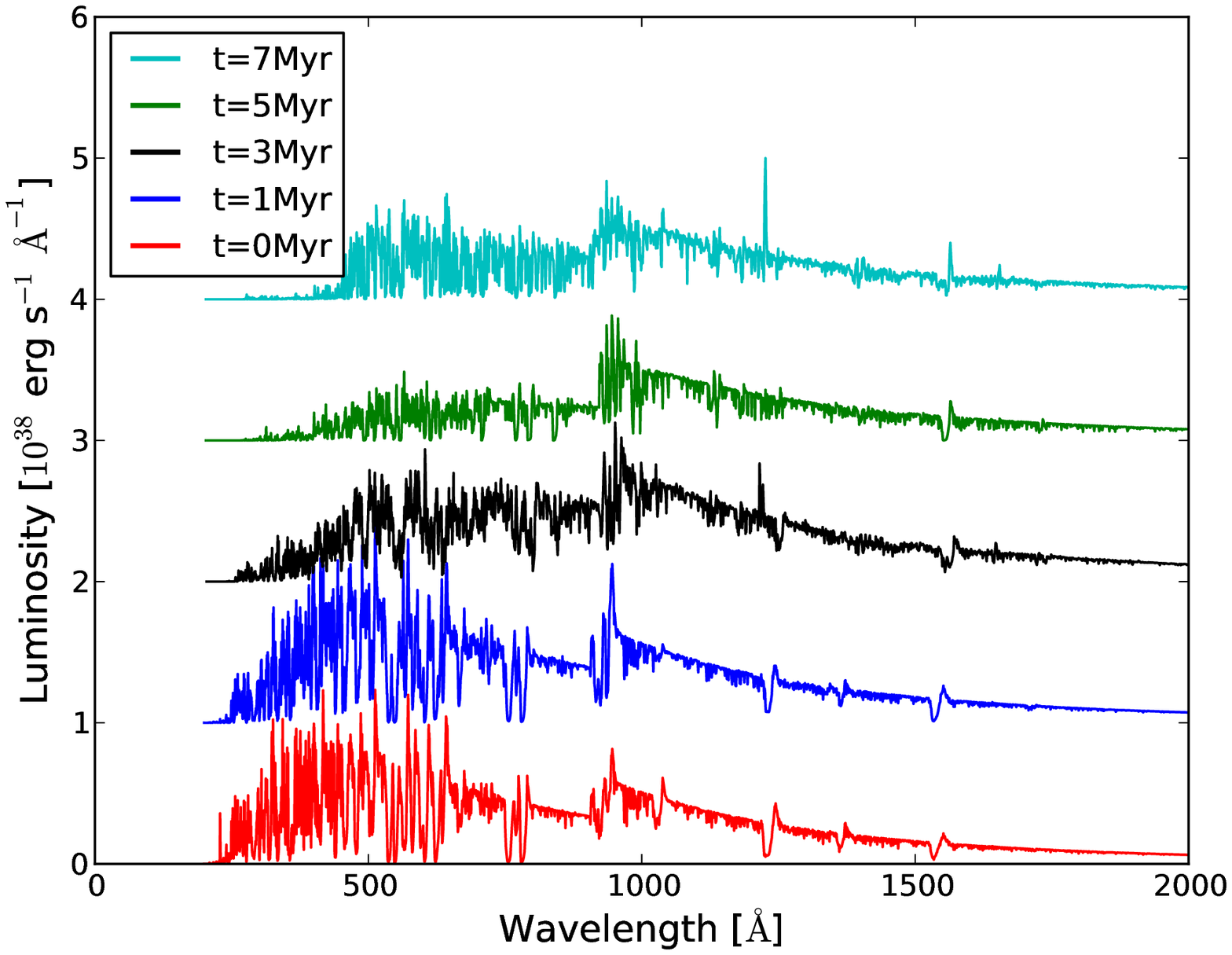}
    \plotone{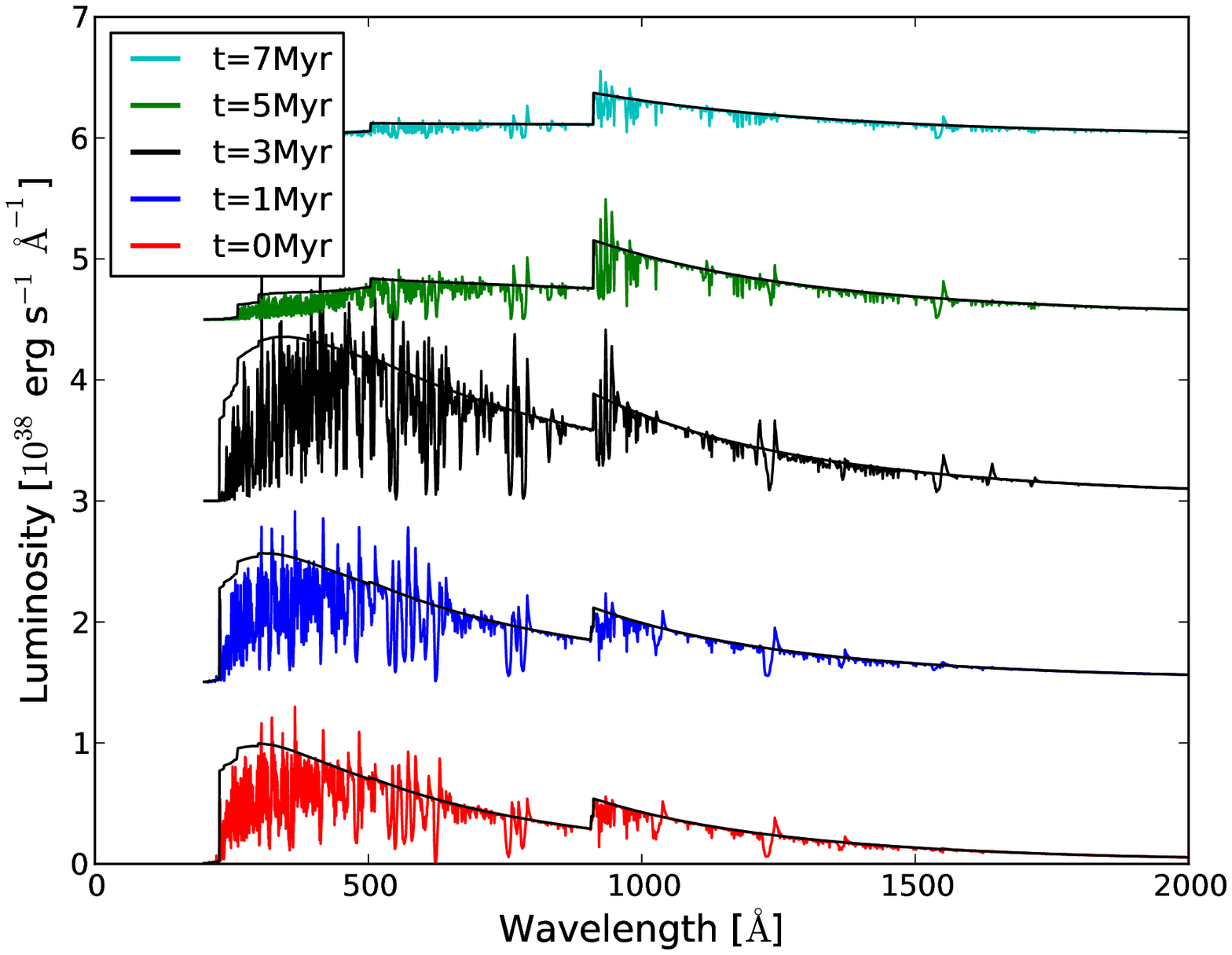}
    \caption{Overlaid spectral distributions in specific luminosity, $L_{\lambda}$ in $10^{38}$
    erg~s$^{-1}$~\AA$^{-1}$, for Monte-Carlo models of a star-forming cluster.  We assume
    $10^5~M_{\odot}$ of stars with Salpeter IMF ($0.1 \leq m \leq 120$) and use evolutionary 
    tracks with rotation and model atmospheres at the appropriate metallicities.  {\it Top:}  
    Solar metallicity ($Z = 0.014$, Ekstr\"om \etal\ 2012).   {\it Bottom:}   Sub-solar metallicity 
    ($Z = 0.002$, Georgy  \etal\ 2013).  Each curve is offset by 1.0 (in $10^{38}$ units) plotted,
    from bottom to top, at times $t =$ 0, 1, 3, 5, and 7 Myr after a coeval burst of star formation.
    See Section 4.5 for details.  }
\end{figure}

%%%%%%%%%%%%%%%%%%%%%%%%%%%%%%%%%%%%%%%%%%%%%%%%%%%%%%%%%%%%%%%

%%%%%%%%%%%%   Figure 13    %%%%%%%%%%%%%%%%%%%%%%%%%%%%%%%%%%%%%%%%%%

\begin{figure} 
   \epsscale{1.2}
     \plotone{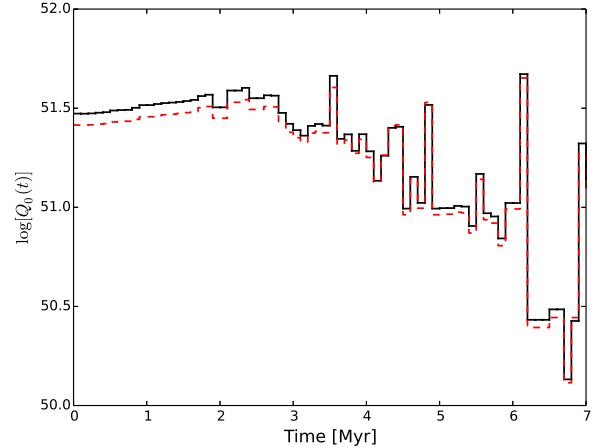}
     \caption{Monte-Carlo simulations, as in Fig.\ 12, of the time evolution of LyC production rate $Q_0(t)$ [photons~s$^{-1}$] 
     in a cluster with $10^5~M_{\odot}$, formed coevally in a Salpeter mass distribution ($0.1 < m < 120$).  We use 
     evolutionary tracks at solar metallicity with rotation ($Z = 0.014$, Ekstr\"om \etal\ 2012) and show two runs with 
     different selections of stellar masses, from initial random-number seeds.   Late-time spikes ($3.5- 6.2$~Myr) arise 
     from the post-main-sequence evolution of stars with 40-60~\msun.  
         }
\end{figure}

%%%%%%%%%%%%%%%%%%%%%%%%%%%%%%%%%%%%%%%%%%%%%%%%%%%%%%%%%%%%%%

The spectra of each model are calculated from the star's basic parameters on the evolutionary tracks, multiplied by the 
number of stars for each mass and added to the total spectra.  The stars are randomly chosen, weighted by the IMF, until 
the total mass of the cluster is $10^5$~\msun.    Owing to the rarity of O-stars at the upper end of the IMF, some models 
had no stars above 100~\msun.    The LyC fluxes do not fade until after 5--7 Myr (Figure~13) owing to their post main-sequence 
evolution to O supergiants.  However, fluxes at the shortest wavelengths ($\lambda < 504$~\AA) decrease after $\sim5$ Myr.  

As noted in Section 3.2, extragalactic observers have calibrated escaping LyC radiation from the flux-density ratios,
$F(912^-) / F(912^+)$ at the Lyman edge, and $F(1500) / F(910)$, the far-UV-to-LyC ratio at 1500~\AA\  and 900~\AA.
These computed flux ratios are listed in Table 3 for model clusters calculated with rotating-star models at solar metallicity 
($Z = 0.014$) and sub-solar metallicity ($Z = 0.002$).   For coeval starburst models, the intrinsic flux decrement at the 912~\AA\
Lyman edge ranges from $F(912^-) / F(912^+) \approx 0.4-0.7$ over the first 5 Myr for both solar and sub-solar metallicities.  
The far-UV (1500~\AA) to LyC (900~\AA) ratio ranges from $F(1500) / F(900) \approx 0.4-0.7$ over the same time period.   
In their studies of escaping LyC from starburst galaxies at $z \approx 3$, Steidel \etal\ (2001) defined a {\it relative} fraction 
of escaping LyC photons normalized to escaping 1500~\AA\ photons:
\begin{equation}
   f_{\rm esc,rel} = \frac { [F(1500)/F(900)]_{\rm int} } { [F(1500)/F(900)]_{\rm obs} } \; \exp( \tau_{\rm IGM,900}) \;  , 
\end{equation}
where $[F(1500) / F(900)]_{\rm int}$ is the intrinsic ratio of non-ionizing, far-UV (1500~\AA) photons to ionizing
photons (at 900~\AA) and $\tau_{\rm IGM,900}$ is the line-of-sight optical depth of the intergalactic medium to
LyC photons at 900~\AA.  The intrinsic spectral shape for massive stars above and below the Lyman edge is
not well constrained by observations, and one must rely on stellar atmosphere models and population synthesis.  
Both Steidel \etal\ (2001) and Inoue \etal\ (2005) adopted an intrinsic ratio $F(1500) / F(910) = 3.0$.   Because 
our (1500~\AA/900~\AA) ratios are considerably lower (0.4-0.5) this will reduce the inferred escape fractions in 
high-redshift starburst  galaxies, which are proportional to the assumed intrinsic ratio.   This decrease in the
intrinsic ratio may resolve the paradoxical inference (Shapley \etal\ 2006), who found a corrected flux ratio, 
$F(1500)/F(900) \approx 2.9$ implying that $f_{\rm esc}$ is greater than 100\%.   After 5--7 Myr, when the most 
massive O-type stars evolve off the main sequence and eventually die, the (1500-to-900~\AA) flux ratios 
begin to increase.  

\section{DISCUSSION AND CONCLUSIONS}

This work provides an update of the production rates of ionizing photons from massive stars, based
on a new generation of stellar evolutionary tracks coupled to modern stellar atmosphere models.   For a range
of parameters (tracks, metallicities, rotation), we find a 50\% increase in the LyC production efficiencies compared
to previous calibrations.  The new stellar tracks from the Geneva group were calculated for both solar and sub-solar 
metallicities, and they also examined the effects of stellar rotation with surface velocities 40\% of breakup.  For the
grid points ($T_{\rm eff}, \log g$) on these tracks, we computed model atmospheres with the non-LTE 
code \texttt{WM-basic}, which treats 3D expanding atmospheres and line-blanketing by lines of heavy elements 
whose metallicities were consistent with those in the stellar tracks.  We computed the production rates of ionizing 
photons in the ionizing continuum of hydrogen, as well as of \HeI\ and \HeII, and their dependence on metallicity, 
rotation, and initial stellar mass.  Next, we integrated these rates over three different IMFs (Salpeter, Kroupa, Chabrier) 
to derive  LyC photon production efficiencies, $Q_{\rm LyC}$, in units of LyC photons per solar mass of star formation.   

Our calculations used the recent grids of evolutionary tracks from the Geneva group, who made specific assumptions 
about stellar mass loss rates and rotation.  These grids could be revised in the future, to account for improved 
parameterization of mass-loss rates, binary evolution, and the upper extent of the IMF.  Recent observations of massive
stars have introduced uncertainties in mass-loss rates, particularly in late (WR) stages (Gr\"afener \& Vink 2013; 
Smith 2014).  Binaries could affect the stellar populations owing to tides, mass transfer, and mergers (de~Mink \etal\
2013).  Some effects of stellar mergers are effectively included in tracks with increased rotation. 
After considering a range of metallicities and the effects of stellar rotation, the main conclusions of our study can be 
summarized as follows:
\begin{enumerate}

\item  By coupling new stellar evolutionary tracks at several metallicities, with and without rotation, to non-LTE model 
    atmospheres, we find a range of LyC photon production efficiencies $Q_{\rm LyC} = [3.1-9.4] \times 10^{60}$ 
    photons per \msun\ of  star formation.  The median LyC production efficiency is $(6\pm2) \times 10^{60}$ LyC 
    photons per \msun, an average increase of 50\% from previous models.  
         
 \item The new LyC efficiency is equivalent to a SFR calibration of $10^{53.3\pm0.2}$ photons~s$^{-1}$ per \msun~yr$^{-1}$.
     With these new rates, the critical SFR density in a clumpy medium can be written
     $\dot{\rho}_{\rm SFR} = (0.012\; M_{\odot}\; {\rm yr}^{-1}\; {\rm Mpc}^{-3}) [(1+z) / 8]^3  (C_{\rm H} / 3) (0.2/ f_{\rm esc}) 
     (0.006 / Q_{\rm LyC}) T^{-0.845}_4$, scaled to fiducial values of IGM clumping factor $C_H = 3$, LyC escape fraction
     $f_{\rm esc} = 0.2$, and electron temperature $T_e = 10^4$~K.  
  
\item The higher LyC efficiencies make it easier to complete reionization by $z \approx 7$ for various physical
    parameters $C_H$, $f_{\rm esc}$, $T_e$ and extrapolations of the galaxy luminosity function to magnitudes
    well below those observed in the HST/WFC3 deep fields.  For LyC efficiency
    $Q_{\rm LyC} = 6 \times 10^{60}$ photons per \msun, reionization is complete by $z = 7$ for
    $(C_H / f_{\rm esc}) \approx 15-30$ and $T_e = (1-2) \times 10^4$~K.  
    
\item We also computed efficiencies for \HeI\ and \HeII\ continua.  For \HeI, we find $Q_{\rm HeI} = 0.6 \times 10^{60}$ photons 
     per \msun\ (non-rotating, solar metallicity) with substantial (60-110\%) enhancements at sub-solar metallicity and with 
     rotation (90\% boosts at solar metallicity and 22-35\% boosts at sub-solar).  For \HeII, we find $Q_{\rm HeII} = 1 \times 10^{56}$ 
     photons per \msun\ (non-rotating, solar metallicity) and a factor of 4.65 higher at low metallicity.  
    
\item Because of the \HI\ opacity in the upper stellar atmosphere, non-LTE models of stars with $T_{\rm eff} < 50,000$~K 
    exhibit 40-60\% intrinsic LyC edges at 912~\AA.  These should be visible with next-generation UV missions 
    that can observe escaping radiation in the rest-frame LyC of low-$z$ starburst galaxies.
    
\item Monte-Carlo simulations of starbursts find ratios of far-UV (1500~\AA) to LyC (900~\AA) flux density 
      $F(1500) / F(900) \approx 0.4-0.5$ over the first 5 Myr, for models with metallicity $Z = 0.014$ and $0.002$.  
      The flux ratio at the Lyman edge $F(912^-) / F(912^+) \approx 0.4-0.7$ over the same  time interval.  
      
\end{enumerate}

Several improvements could be made in future calculations.   Additional stellar tracks at initial masses between 
25 and 60~\msun, with rotation and super-solar metallicity would strengthen our results for the metallicity dependence 
of $Q_{\rm LyC}$.  They would also provide a further basis of comparison to 
older super-solar evolutionary tracks (Schaller \etal\ 1992).   Models with a range of rotation rates would be helpful, since 
rotation speeds at 40\% of breakup may be excessive and produce ionizing SEDs inconsistent with nebular emission lines
(Levesque \etal\ 2012).  Our calculations rely on rotating tracks in which all stars have the same initial rotation.  None of the 
current tracks include a description of magnetic fields, although that has not yet been shown to have a significant 
effect.    Future stellar population models will treat  rotation more realistically, with better understanding 
of their ionizing photon budgets.  \\

Several changes in our computational scheme could also result in more accurate photon fluxes and production rates.  When 
creating the grid ($\log g$, $T_{\rm eff}$) of model atmospheres, we continued adding models until we reached convergence 
in photon production rates.   Adding more model atmospheres, up to one per time step of the evolutionary tracks, 
would create more precision.  Another small effect is the contribution of ionizing photon production of lower mass stars.  We 
currently consider models with masses $m$ $\ge 25$ \msun.   Lower mass stars produce an order of magnitude or more fewer 
LyC photons.  Filling in the grid at other metallicities and for additional masses between 35~\msun and 120~\msun\ would be
helpful.  For reionization, the most important factors that could increase the total LyC production are the stellar populations
at low metallicity, the extent of the upper main sequence,  the influence of rapid rotation, and the role of binaries. \\

%%%%%%%%%%%%%%%%%%%%%%%%%%%%%%%%%%%%%%%%%%%%%%%%%%%%%%%%%%

\acknowledgments

\noindent
We thank Claus Leitherer, Emily Levesque, Phil Massey, Max Pettini, John Stocke, and Evan Tilton for helpful discussions 
about stellar atmosphere models, photoionizing backgrounds, and Wolf-Rayet populations, and Anthony Harness for updating 
our cosmological reionization simulator.  We thank the referee for emphasizing the uncertainties in the LyC production 
calculations arising from effects of stellar mass-loss, Wolf-Rayet stars, and binary interactions. 
This research was supported by the STScI COS grant (NNX08-AC14G) at the University of Colorado Boulder.   JMS thanks 
the Institute of Astronomy at Cambridge University for their stimulating scientific environment and support through the 
Sackler Visitor Program.

%%%%%%%%%%%%%%%%%%%%%%%

 \clearpage

%%%%%%%%%%%%   Table 1   %%%%%%%%%%%%%%%%%%%%%%%%%%%%%%%%%%%%%

\begin{deluxetable}{lccccccc}
\tabletypesize{\footnotesize}
\tablecaption{Ionizing Photon Production\tablenotemark{a}  vs.\  Initial Stellar Mass  } 
\tablecolumns{8}
\tablewidth{0pt}

\tablehead{ \colhead{Model Tracks\tablenotemark{b}} &  \colhead{$(Z/Z_{\odot})$ }  & 
 \colhead{$Q_0(120~M_{\odot})$}  & \colhead{$Q_0(85~M_{\odot})$} &  \colhead{$Q_0(60~M_{\odot}$)} &  
 \colhead{$Q_0(40~M_{\odot})$} &  \colhead{$Q_0(32~M_{\odot})$} &  \colhead{$Q_0(25~M_{\odot})$} \\
  &  & ($10^{63}$~phot) & ($10^{63}$~phot) & ($10^{63}$~phot) & ($10^{63}$~phot) 
  & ($10^{63}$~phot) & ($10^{63}$~phot)  } 

\startdata
Ekstr\"om (Rot)           & 0.014  & 11.5  & 9.34 & 6.32  & 3.19 & 2.06   & 1.03   \\
Ekstr\"om (Non-Rot)  & 0.014  & 7.49  & 5.11 & 3.58  & 1.58 & 0.902 & 0.445 \\
Georgy (Rot)               & 0.004  & 16.4  & 9.95 & 5.70  & 2.59 & 1.76    & 1.67   \\  
Georgy (Non-Rot)      & 0.004  & 9.95  & 7.31 & 4.47  & 2.19 & 1.43    & 0.871 \\
Schaller (Non-Rot)    & 0.020   & 7.91 & 5.78 & 3.66  & 2.42 & \dots    & 0.474 \\
Schaller (Non-Rot)    & 0.004  & 8.29  & 6.46 & 4.47  & 2.18 & \dots    & 0.648  
\enddata

 \tablenotetext{a} {Total number of LyC photons, $Q_0(m)$ in $10^{63}$ photons, produced over 
 stellar lifetime for stars of initial mass $m$.  Evolutionary tracks and non-LTE model atmospheres are 
 described in Section 3 and Table 2 footnote.}  
 
\tablenotetext{b} {Evolutionary track references:  Schaller \etal\  (1992); Ekstr\"om \etal\ (2012); Georgy \etal\ (2013),
 of various metallicities ($Z/Z_{\odot})$ and rotating or non-rotating models as noted.  }

\end{deluxetable}

%%%%%%%%%%%%%%%%%%%%%%%%%%%%%%%%%%%%%%%%%%%%%%%%%%%%%%%%%%%%%

%%%%%%%%%%%%  Table 2   %%%%%%%%%%%%%%%%%%%%%%%%%%%%%%%%

\begin{deluxetable}{lcclccccccc}
\tabletypesize{\footnotesize}
\tablecaption{Ionizing Photons\tablenotemark{a} with Various IMFs, Tracks,  and Metallicities ($Z$)  } 
\tablecolumns{11}
\tablewidth{0pt}

\tablehead{ \colhead{Model}  &  \colhead{IMF\tablenotemark{b}} & \colhead{Tracks\tablenotemark{b}} 
    & \colhead{Rot-Type} &  \colhead{$m_{\rm min}$}  & \colhead{$m_{\rm max}$} &  \colhead{Slope} 
    &  \colhead{Metals} & \colhead{$Q_{\rm LyC}$} & \colhead{$Q_{\rm HeI}$} & \colhead{$Q_{\rm HeII}$}  \\   
    &   &    &    &  (\msun)  &  (\msun)  &  $\alpha$ & $Z$ & ($10^{61}$) & ($10^{60}$) &  ($10^{56}$)      } 

\startdata
Solar  &  1   & 4  & Nonrot     &    $0.1$    &  $120$    &   $2.35$  &  $0.020$   & $0.367$   & 0.552  &  0.918  \\
Solar  &  1   & 5  &    Rot        &    $0.1$    &  $120$    &   $2.35$  &  $0.014$   & $0.603$   & 1.11    &   1.73 \\
Solar  &  1   & 5  & Nonrot     &    $0.1$    &  $120$    &   $2.35$  &  $0.014$   & $0.322$   & 0.571  &  1.17  \\
\hline
Solar  &  2   & 4  & Nonrot     &    $0.01$  &  $120$    &   $2.3 $   &  $0.020$   & $0.572$   & 0.864 & 1.45   \\
Solar  &  2   & 5  &   Rot         &    $0.01$  &  $120$    &   $2.3 $   &  $0.014$   & $0.940$   &  1.74   &  2.76  \\
Solar  &  2   & 5  & Nonrot     &    $0.01$  &  $120$    &   $2.3 $   &  $0.014$   & $0.503$   &  0.895 &  1.86 \\
\hline
Solar  &  3   & 4  & Nonrot     &    $0.01$  &  $120$    &   $2.3 $   &  $0.020$   & $0.370$   & 0.558  &  0.935  \\
Solar  &  3   & 5  &  Rot          &    $0.01$  &  $120$    &   $2.3 $   &  $0.014$   & $0.607$   & 1.13    &  1.78  \\
Solar  &  3   & 5  & Nonrot     &    $0.01$  &  $120$    &   $2.3 $   &  $0.014$   & $0.325$   & 0.578  &  1.21  \\

\hline\hline \\

Sub-solar & 1   & 4  & Nonrot     &    $0.1$    &  $120$    &   $2.35$  &  $0.004$  & $0.399$ & 0.884 & 4.25   \\
Sub-solar & 1   & 6  &    Rot        &    $0.1$    &  $120$    &   $2.35$  &  $0.002$  & $0.583$ & 1.47   &  8.08   \\
Sub-solar & 1   & 6  & Nonrot     &    $0.1$    &  $120$    &   $2.35$  &  $0.002$  & $0.433$  & 1.21  &  10.7    \\
\hline
Sub-solar & 2   & 4  & Nonrot     &    $0.01$  &  $120$    &   $2.3 $   &  $0.004$  & $0.623$ & 1.38 & 6.75  \\
Sub-solar & 2   & 6  &    Rot        &    $0.01$  &  $120$    &   $2.3 $   &  $0.002$  & $0.910$ & 2.31 & 12.8   \\
Sub-solar & 2   & 6  & Nonrot     &    $0.01$  &  $120$    &   $2.3 $   &  $0.002$  & $0.676$ & 1.90 & 16.9    \\
\hline
Sub-solar & 3  & 4  & Nonrot     &    $0.01$  &  $120$    &   $2.3 $   &  $0.004$  & $0.402$ & 0.896 &  4.36   \\
Sub-solar & 3   & 6  &    Rot        &    $0.01$  &  $120$    &   $2.3 $   &  $0.002$  & $0.588$& 1.49   &  8.28     \\
Sub-solar & 3   & 6  & Nonrot     &    $0.01$  &  $120$    &   $2.3 $   &  $0.002$  & $0.436$ & 1.22 &   10.9    
\enddata

\tablenotetext{a}{Photon production,$Q_{\rm LyC}$, $Q_{\rm HeI}$, $Q_{\rm HeII}$ in last three columns
are for ionizing continua of \HI, \HeI, and \HeII, respectively.   Total LyC production ($Q_{\rm LyC}$) 
is expressed in units of $10^{61}$ lifetime-integrated photons per \msun\ of star formation, calculated for 
various IMFs and metallicities (columns 1-8) and for evolutionary tracks with or without rotation.   Values
of $Q_{\rm HeI}$ and $Q_{\rm HeII}$ are in units of $10^{60}$ and $10^{56}$ photons per \msun, as noted 
in header.}

\tablenotetext{b}{IMF references:  (1) Salpeter (1955); (2) Kroupa (2001); (3) Chabrier (2003). 
Evolutionary track references:  (4) Schaller \etal\  (1992); (5) Ekstr\"om \etal\ (2012); (6) Georgy \etal\ (2013). 
Using the Kroupa or Chabrier IMFs increases the LyC production relative to the Salpeter IMF by 
factors of 1.56 and 1.64, respectively.  
} 

\end{deluxetable}

%%%%%%%%%%%%%%%%%%%%%%%%%%%%%%%%%%%%%%%%%%%%%%%%%%%%

%%%%%%%%%%%%   Table 3   %%%%%%%%%%%%%%%%%%%%%%%%%%%%%%%%%%%%%

\begin{deluxetable}{ccccc}
\tabletypesize{\footnotesize}
\tablecaption{Flux Ratios in Composite Cluster Spectra\tablenotemark{a} } 
\tablecolumns{5}
\tablewidth{0pt}

\tablehead{ \colhead{Time\tablenotemark{b}} & \colhead{ $F(912^-) / F(912^+)$} & \colhead{ $F(912^-) / F(912^+)$}  &
                      \colhead{ $F(1500) / F(900)$} &  \colhead{ $F(1500) / F(900)$}    \\
                 &  \colhead{ $Z = 0.014$} & \colhead{ $Z = 0.002$} & \colhead{ $Z = 0.014$} & \colhead{ $Z = 0.002$} }

\startdata
0~{\rm Myr} &  0.54  &  0.58  &  0.472  &  0.455  \\
1~{\rm Myr} &  0.57  &  0.71  &  0.457  &  0.438  \\
3~{\rm Myr} &  0.65  &  0.71  &  0.506  &  0.404  \\
5~{\rm Myr} &  0.39  &  0.41  &  0.777  &  0.727  \\
7~{\rm Myr} &  0.50  &  0.31  &  0.630  &  1.03  
\enddata

 \tablenotetext{a}{Ratios of flux densities, $F_{\lambda}$ (erg~cm$^{-2}$ s$^{-1}$~\AA$^{-1}$) at key wavelengths:  
    $F(912^-)/F(912^+)$ is the flux decrement at the 912~\AA\ Lyman edge; $F(1500)/F(900)$ is the ratio of fluxes 
    in the far UV (1500~\AA) and 900~\AA, just shortward of the LyC edge (Steidel \etal\ 2001;  Shapley \etal\ 2006).  
    These ratios are shown for simulated clusters (Figure 12) with solar and sub-solar metallicities, as noted, using
    evolutionary tracks with rotation (Ekstr\"om \etal\ 2012; Georgy \etal\ 2013).  }
 
 \tablenotetext{b} {Time in Myr following a coeval burst of star formation in a cluster with $10^5$~\msun\ with
  a Salpeter IMF ($0.1 < m < 120)$ as shown in Figure 12. }  

\end{deluxetable}

%%%%%%%%%%%%%%%%%%%%%%%%%%%%%%%%%%%%%%%%%%%%%%%%%%%%%%%%%%%%%

\end{document}